\documentclass[12pt, hyper]{article}
\usepackage{color}
\usepackage[table]{xcolor}
\usepackage{amsmath}
\usepackage{amsfonts}
\usepackage{amsopn}
\usepackage{graphicx} 
\usepackage{adjustbox}
\usepackage{amssymb}
\usepackage{amsthm}
\usepackage{hyperref}
\usepackage{tikz-cd}
\usepackage{dynkin-diagrams}
\usepackage{mathrsfs}
\DeclareMathAlphabet{\mathpzc}{OT1}{pzc}{m}{it}
\usepackage{verbatim}
\usepackage{multirow}
\usepackage{enumitem}
\usepackage{setspace} 
\usepackage{arydshln}
\usepackage{cite}
\usepackage{amsthm}
\usepackage{cleveref}
\usepackage{cite}
\usepackage{amsthm}
\usepackage[margin=1cm]{caption}
\usepackage[font={small}]{caption}

\def\CH{{\cal H}}
\def\CM{{\cal M}}

\def\CS{{\cal S}}
\def\CT{{\cal T}}

\def\a{\alpha}\def\b{\beta}\def\g{\gamma}
\def\d{\delta}
\def\th{\theta}
\def\l{\lambda}
\def\n{\nu}
\def\r{\rho}
\def\t{\tau}
\def\G{\Gamma}

\def\be{{\bar e}}

\def\be{\begin{equation}}
\def\ee{\end{equation}}

\newtheorem*{conj}{Conjecture}
\newtheorem{thm}{Theorem}[section]

\usepackage{cleveref}

\begin{document}

\begin{titlepage}
\vfill
\begin{flushright}
{\tt\normalsize KIAS-P22065}\\

\end{flushright}
\vfill
\begin{center}
{\Large\bf On Classification of Fermionic \\ \mbox{}\vspace{-5mm} \\
Rational Conformal Field Theories }

\vskip 1cm

{  Zhihao Duan, Kimyeong Lee,  Sungjay Lee and Linfeng Li}

\vskip 5mm
\vskip 3mm {\it Korea Institute for Advanced Study \\
$~~ $85 Hoegiro, Dongdaemun-Gu, Seoul 02455, Korea}
\end{center}
\vfill

\begin{abstract}
\noindent

We systematically study how the integrality of 
the conformal characters shapes the space of 
fermionic rational conformal field theories in two dimensions. 
The integrality suggests that
conformal characters on torus with a given choice of spin structures should be invariant under a principal congruence subgroup of $\mathrm{PSL}(2,\mathbb{Z})$.
The invariance strongly constrains the possible values of the central charge as well as the conformal weights in both Neveu-Schwarz and Ramond sectors, which improves the conventional holomorphic modular bootstrap method in a significant manner. This allows us to make much progress on the classification of fermionic rational conformal field theories with the number of independent characters less than five. 

\end{abstract}

\vfill
\end{titlepage}

\renewcommand{\thefootnote}{\#\arabic{footnote}}
\setcounter{footnote}{0}

\tableofcontents

\section{Introduction and Conclusion}

Conformal field theories (CFTs) play prominent roles in theoretical physics ranging from the critical phenomena of phase transitions, the boundary excitation of (fractional) quantum Hall effects, to the world-sheet dynamics of quantum strings. If we specialize to two dimensions, the conformal symmetry gets enlarged to the infinite dimensional Virasoro algebra 
that restricts the underlying dynamics severely. As a consequence, for example,  any unitary CFT with central charge $c$ less than one can only be one of the minimal models\cite{Belavin:1984vu}.

However, for general 2d CFTs, it still remains a difficult task to comprehend  the  full landscape of their theory space.   After the breakthrough in the study of 3d Ising model \cite{PhysRevD.86.025022}, the philosophy of bootstrap has taken a more central role in recent years. In particular, it became more compelling  to explore  the idea of modular bootstrap, which  employs  the modular invariance of the CFT partition functions on the torus. The modular invariance  implies new constraints on the space of 2d CFTs as well as possible 3d gravity duals, and also provides unexpected connections to mathematics. See for example \cite{Hellerman:2009bu,Friedan:2013cba,Hartman:2014oaa,Bae:2017kcl,Bae:2018qym,Benjamin:2019stq,Hartman:2019pcd,Afkhami-Jeddi:2020ezh,Meruliya:2021utr,Lin:2021udi,Grigoletto:2021zyv,Benjamin:2021ygh,Benjamin:2022pnx} for a very partial list of related works.

Rational conformal field theories (RCFTs), defined to have finitely many chiral primaries, have drawn particular attention in the past decades. The minimal models, lattice CFTs, and the Wess-Zumino-Witten (WZW) models of compact group are just a few examples of the RCFT. Compared to irrational CFTs, they have much nicer properties, and we are naturally led to a quest for its possible classification. 

Mathur, Mukhi and Sen \cite{Mathur:1988na} first realized that modular symmetry can be explored to systematically classify bosonic RCFTs, based on the number of independant characters $d$, the \emph{rank} of RCFT, they have. This approach is dubbed as holomorphic modular bootstrap \cite{Chandra:2018pjq,Mukhi:2019xjy,Mukhi:2022bte}. Its essential idea is the observation that characters form a vector-valued modular function (vvmf), and mathematically they must satisfy a modular linear differential equation (MLDE). We also remark that the method of MLDE also features in the study of higher dimensional quantum field theories, most notably in four-dimensional superconformal field theories (SCFTs) through the so-called SCFT/VOA correspondence \cite{Beem:2013sza,Beem_2018,Pan:2021ulr,Kaidi:2022sng,Zheng:2022zkm,Hatsuda:2022xdv}.   

On the other hand, in mathematics there is a class of different although related objects: modular tensor categories (MTCs). RCFTs and MTCs are similar because modularity  plays important role in both cases. For instance the modular data of an MTC should in some sense capture the modular transformation of RCFT characters. The classification of MTCs according to their rank is also an important topic \cite{rowell2009classification,HONG20101000,bruillard2016classification}, which resonates with the holomorphic modular bootstrap method. However, since it is still unknown whether every MTC is realized by an RCFT, and even if so a given MTC could be mapped to many RCFTs, we are still lacking a precise dictionary between the two methods.

Nevertheless, this does not stop us from utilizing techniques developed in the MTC side. In particular, the so-called congruence property \cite{Ng:2012ty} is utilized to constrain the set of modular data for MTC at low ranks. The latest result can be found in \cite{ng2022reconstruction}. In the RCFT side, this can be formulated as the integrality conjecture or unbounded denominator conjecture \cite{atkin1971modular}, which was recently proved in \cite{calegari2021unbounded} (see also \cite{bantay2007vectorvalued}). Its statement is that each component of a vvmf becomes a modular function for a congruence subgroup $\Gamma(N)$ of $\mathrm{SL}(2,\mathbb{Z})$ if all the coefficients in its $q$-expansion are integral. Last year, \cite{Kaidi:2021ent} imported this technique to RCFTs and greatly extended the previous classification using holomorphic modular bootstrap.

In this paper, we will consider a generalization of the above successful story to theories including fermions.  This generalization was initiated in the papers \cite{Bae:2020xzl,Bae:2021mej}, in which one examines the modular subgroups preserved by the choice of spin structure for fermions, and write down corresponding fermionic MLDEs (FMLDEs). Naturally, this tool helps to classify what one may call fermionic RCFTs (FRCFTs), which has interesting connections to various topics such as fermionization \cite{Gaiotto:2015zta,Karch:2019lnn}, emergence of supersymmetry (SUSY) \cite{Bae:2021lvk,Kikuchi:2021qxz,Bae:2021jkc,Kikuchi:2022jbl}, moonshine phenomena of sporadic groups \cite{Bae:2020pvv,Bae:2021mej}, etc.

As a next step, naturally we would like to study the implication of integrality of the Fourier coefficients of the characters for FRCFTs. We state an analogue of the integrality conjecture in \ref{Sec:Integrality}, which is the counterpart of congruence property in super-MTC \cite{bonderson2018congruence}. 
Assuming this conjecture, we are able to, in certain sense, extend the previous classification in \cite{Bae:2020xzl,Bae:2021mej}, and we successfully bootstrap candidate solutions for putative FRCFT characters up to rank four. To be more specific, first of all, by our working hypothesis, we are able to cover {\it non-degenerate} FRCFTs, meaning that there exists no pair of NS sector conformal weights whose difference is a multiple of a half-integer, and no pair of R sector conformal weights whose difference is an integer. For this class, then indeed all the theories found in \cite{Bae:2020xzl,Bae:2021mej} are recovered. Moreover, there is a non-negative integer $\ell$ as another input parameter, which characterizes the pole structure of the coefficient functions of FMLDE and is known as the index. We only consider index $\ell \leq 1$ in this paper, together with possible unitarity constraint. For easy of reference, we summarize the main results here:
%\vspace{0.5mm}
\begin{flushleft}
\begin{tabular}{cccc}
     $(d, \ell) = (2,0)$
     & $(d, \ell) = (3,0)$\, \text{unitary} & $(d, \ell) = (3,1)\, \text{unitary}$ 
     & $(d, \ell) = (4,0)$\, \text{unitary}\\[5pt]
     Tables \ref{tab:rktwo1}, \ref{tab:rktwo2}
    &   Table \ref{Tab:rank3l0}
     & Table \ref{Tab:rank3l1} & Table \ref{tab:rk4l0} 
\end{tabular}
\end{flushleft}
\vspace{2mm}

We would like to stress here that the previous approaches to the  classification of the FRCFTs as well as some studies in the super-MTC are limited in a sense that they mainly rely on the physical constraints in the Neveu-Schwarz (NS) sector but barely concern those in the Ramond (R) sector. There are however some occasions where the torus partition function in the NS sector,  that looks perfectly consistent, is modular-transformed to the partition function in the R sector that is ill-defined. Recently, it was pointed out in \cite{Benjamin:2020zbs} that 
careful examining the often-ignored Ramond sector results in a stronger constraint  on the spectra of fermionic CFTs. Interestingly, we observe that the consequence of the integrality conjecture for FRCFTs actually not only constrains the spectra in both NS and R sectors but also the provides a consistency relation between them. 

For readers who wish to compare our results with the classification in the super-MTC literature, we first remark that the exponential of conformal weights in their language are known as twists or topological spins, while their central charge $c$ is only defined mod 8. Also, to obtain their normalization of $T$ matrix in the modular data, one needs to multiply ours by an overall factor $\exp(c/24)$. Therefore, the number $N$ which labels the congruence subgroup will in general differ from those appearing in the modular data. More importantly, due to the fact that different primaries may share the same character, the rank of super-MTC will in general be bigger than the number of independent characters in our classification. 

As possible further directions, first it would be nice to understand or disprove the solutions that we find but are unable to identify. 
For instance, one has to check if they have the well-defined fusion algebra.  
Although each fusion coefficient can be computed by the Verlinde formula, 
it requires the so-called refined modular matrices.   
However, the conformal characters constructed by the holomorphic modular bootstrap 
only provide the reduced modular matrices that leads to the wrong fusion coefficients.
See \cite{Mathur:1988gt} for reference. 
It would be interesting to develop a systematic manner to unfurl the reduced 
modular matrices to refined modular matrices that eventually furnishes the (super) MTC data, whose classification can be found in \cite{bruillard2019classification,bruillard2020classification}. 
Second, one could consider turning on extra parameters such as flavor fugacity in the characters, which upgrades the MLDEs to the so-called flavored MLDEs. Such generalization has already appeared, for example, in \cite{Pan:2021ulr,Zheng:2022zkm}. Third, it would be interesting to look for a correct Hecke operator relating different FRCTs, as was done for bosonic RCFTs in \cite{Harvey:2018rdc,Duan:2022ltz}.
Finally, incorporating all possible topological defect lines for fermionic theories would be another important direction to pursue \cite{Chang:2022hud}, and FRCFTs are surely suitable examples to study.

This paper is organized as follows. In section \ref{sec:2}, we review the basic structure of 2d CFT in the presence of fermions, and introduce the holomorphic modular bootstrap method. New ingredients start from section \ref{Sec:Integrality} where we make use of the integrality conjecture and hence the representation theory of finite groups to constrain all possible exponents mod 1 for fermionic MLDEs at low ranks. Section \ref{sec:3} contains the main result of this paper, where we present explicitly putative two-, three- and four-character fermionic theories with constraint mentioned above after a computerized scan. In particular, they contain all non-degenerate theories previously found in \cite{Bae:2020xzl,Bae:2021mej}. We also include two appendices. In Appendix \ref{App:Group}, we give some detail about the induced representation to prove a claim in Section \ref{Sec:Integrality}. In Appendix \ref{App:exponents}, we list all the exponents mod 1 that are used in Section \ref{sec:3}.

{\it Note Added:} While this work is at the final stage, \cite{Cho:2022kzf} appeared which 
uses the same idea to construct modular data in the super-MTC.

%%%%%%%%%%%%%%%%%%%%%%%%%%%%%%%%%%%%%%%%%%%%%%%%%%%%%%%%%%%%%%%%%%%%%%%%%%%%%%%%%%%%%%%%%%%%%%%%%%%%%%%%%%%%%%%%
%%%%%%%%%%%%%%%%%%%%%%%%%%%%%%%%%%%%%%%%%%%%%%%%%%%%%%%%%%%%%%%%%%%%%%%%%%%%%%%%%%%%%%%%%%%%%%%%%%%%%%%%%%%%%%%%

\section{Preliminaries}\label{sec:2}

An RCFT is defined as a CFT whose torus partition function 
can be described as a finite sum of products of  holomorphic functions and anti-holomorphic 
functions of the complex parameter of torus $\tau$,
\begin{align}
    Z(\tau,\bar\tau)={\rm Tr}_{\mathcal{H}} \big[q^{L_0-c/24} \bar q^{\bar L_0-c/24}\Big]= \sum_{i,j=0}^{d-1} M_{ij}\chi_i(\tau)\bar\chi_j(\bar\tau)
\end{align}
with the trace over the Hilbert space $\mathcal{H}$ on a circle and  non-negative integers $M_{ij}$. Such holomorphic functions $\chi_i(\tau)$ with $i=0,1,...,d-1$ are referred to as conformal characters of the chiral primary operators with respect to a 
certain chiral algebra that includes the Virasoro algebra. The central charge $c$ and the conformal weights $h$ of the primary operators turn out be rational numbers for RCFTs.

The invariance of the partition function of a bosonic RCFT 
under the modular transformation of torus, $\mathrm{SL}(2,\mathbb{Z})$,  leads 
to the fact that conformal characters, denoted by $\vec {\chi}(q)$ ($q=e^{2\pi i \tau}$) collectively, 
transform as $d$-dimensional vector-valued modular forms of weight zero. For instance, 
the character $\chi_i(q)$ transform under $S$ and $T$ as follows
\begin{align}
    \chi_i(-1/\tau) = \sum_{j=0}^{d-1} \CS_{ij}\chi_i(\tau) , 
    \qquad 
    \chi_i(\tau+1) = \sum_{j=0}^{d-1} \CT_{ij}\chi_i(\tau), 
\end{align}
where the modular matrices $\CS$ and $\CT$ are constant matrices in $\mathrm{GL}(d,\mathbb{C})$, and satisfy 
the relations  
\begin{align}
    \CS^2 = (\CS\CT)^3 = C\     
\end{align}
with $C$ the charge conjugation matrix.

\begin{figure}[t!]
    \centering
    \includegraphics[width=13cm]{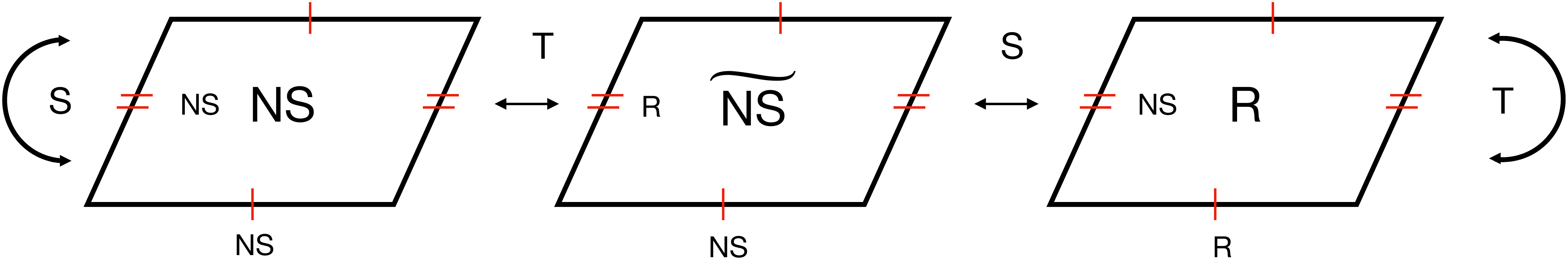}
    \caption{Transformations among NS, $\widetilde{\text{NS}}$ and R sectors.}
    \label{fig:threesectors}
\end{figure}
In the presence of fermions, the boundary condition along the non-trivial 
cycles relaxes the $\mathrm{SL}(2,\mathbb{Z})$ invariance of the torus partition function. To see this, note that one can impose either periodic (R) or anti-periodic (NS)
boundary condition for a fermion along each cycle of torus.  As 
depicted in Fig. \ref{fig:threesectors}, we 
thus have four possible boundary conditions,  
labelled as (NS,NS), (R,NS), (NS,R), and (R,R), to define a theory of 
fermions on torus.  A choice of boundary conditions is also known as 
a choice of spin structures for the fermions. 
For convenience, in this work we use a shorthand notation 
NS, $\widetilde{\text{NS}}$, R and $\widetilde{\text{R}}$ for the aforementioned 
boundary conditions,  respectively. For the interacting field theories, we impose the same boundary conditions on all fermions.
The torus partition function for each boundary condition 
allows the Hamiltonian interpretation as follows, 
\begin{align} \label{Hamiltonian01}
  Z_\text{NS} (\tau,\bar \tau) & = \text{Tr}_{\CH_\text{NS}} \Big[ q^{L_0-c/24} {\bar q}^{\bar L_0 -c/24} \Big],
  \nonumber \\
  Z_{\widetilde{\text{NS}}} (\tau,\bar \tau) & = 
  \text{Tr}_{\CH_\text{NS}} \Big[(-1)^F q^{L_0-c/24} {\bar q}^{\bar L_0 -c/24} \Big],
  \\
  Z_\text{R} (\tau,\bar \tau) & = \text{Tr}_{\CH_\text{R}} \Big[ q^{L_0-c/24} {\bar q}^{\bar L_0 -c/24} \Big],
  \nonumber \\ 
  Z_{\widetilde{\text{R}}} (\tau,\bar \tau) &= \text{Tr}_{\CH_\text{R}} \Big[(-1)^F q^{L_0-c/24} {\bar q}^{\bar L_0 -c/24} \Big],
  \nonumber
\end{align}
where the trace is performed over the Hilbert space of a given CFT on circle 
in the Neveu-Schwarz (Ramond) sector, $\CH_\text{NS}$ ($\CH_\text{R}$).

We also describe in Fig. \ref{fig:threesectors} how 
the fermionic spin structures transform under $S$ and $T$. It is then 
evident that the torus partition functions for NS, $\widetilde{\text{NS}}$, R  boundary conditions are invariant under $\Gamma_\theta$, $\Gamma^0(2)$, and $\Gamma_0(2)$, 
\begin{gather}
    \Gamma_\theta =  
    \Bigg\{\begin{pmatrix}
        \a& \b\\ \g& \d
    \end{pmatrix} \in \text{SL}(2,\mathbb{Z}), \quad  
    \begin{pmatrix}
        \a& \b\\ \g& \d
    \end{pmatrix} 
    \equiv 
    \begin{pmatrix}
        1&0\\ 0&1
    \end{pmatrix} \ {\rm or}\ 
    \begin{pmatrix}
        0&1\\ 1&0
    \end{pmatrix} 
    \text{mod}\ 2\Bigg\}, \nonumber
\\
    \Gamma^0(2)=  \Bigg\{\begin{pmatrix}
        \a& \b\\ \g& \d
\end{pmatrix} \in \text{SL}(2,\mathbb{Z}),\ \  \b \equiv 0 \  \text{mod}\ 2\Bigg\},\\
    \Gamma_0(2)=  \Bigg\{\begin{pmatrix}
        \a& \b\\ \g& \d
\end{pmatrix} \in \text{SL}(2,\mathbb{Z}),\ \  \g \equiv 0 \  \text{mod}\ 2\Bigg\}, \nonumber
\end{gather}
respectively. They are the level-two congruence subgroups of $\mathrm{SL}(2,\mathbb{Z})$.
On the other hand, the $\mathrm{SL}(2,\mathbb{Z})$ invariance of 
the partition function for the $\widetilde{\text{R}}$ boundary 
condition remains intact.

We refer an FRCFT as a conformal field theory of fermions 
whose partition function for each boundary condition can be 
expressed in terms of finite numbers of conformal characters in each sector, 
\begin{gather}
  Z_\text{NS}  = 
  \sum_{i,j=0}^{d-1} M_{ij} \chi_i^\text{NS}(\tau) \bar{\chi}_j^\text{NS}(\bar \tau), \qquad
  Z_{\widetilde{\text{NS}}}  =  
  \sum_{i,j=0}^{d-1} \widetilde{M}_{ij} \chi_i^{\widetilde{\text{NS}}} (\tau) \bar{\chi}_j^{\widetilde{\text{NS}}}(\bar \tau),
  \nonumber
  \\
  Z_\text{R}  = 
  \sum_{i,j=0}^{d-1} N_{ij}  \chi_i^\text{R}(\tau) \bar{\chi}_j^\text{R}(\bar \tau), \qquad
  Z_{\widetilde{\text{R}}}  =  
  \sum_{i,j=0}^{\tilde{d}-1} \widetilde{N}_{ij} \chi_i^{\widetilde{\text{R}}} (\tau) {\bar \chi}_j^{\widetilde{\text{R}}} (\bar \tau) ,
\end{gather}
where the modular pairing matrices $M_{ij}$, $\widetilde{M}_{ij}$, 
$N_{ij}$, and $\widetilde{N}_{ij}$ are constant matrices with integer 
components. We define the number of independent  characters $d$ to be the rank of FRCFT. The rank $d$ remains the same in $\mathrm{NS}$, $\widetilde{\mathrm{NS}},\mathrm{R}$ as the characters of these three sectors are related by the $T,S$ transformations.  We could treat the $\widetilde{\text{R}}$ sector partition function somewhat like a bosonic RCFTs  as they are $\mathrm{SL}(2,\mathbb{Z})$-invariant. The rank $\tilde{d}$ of $\widetilde{\mathrm{R}}$ sector characters   is independent from $d$.   We would hardly explore the $\tilde{R}$ sector physics in this work. In the limit $\t \to i \infty$, such characters in each sector  can be expanded in powers of $q$ as follows,
\begin{align}\label{exponents}
  \chi^\text{NS}(q) & = q^{\a^\text{NS}} \big( a_0 + a_{1/2} q^{1/2} + a_{1} q^1 + a_{3/2}q^{3/2} + \cdots \big), 
  \nonumber \\ 
  \chi^{\widetilde{\text{NS}}} (q) & = q^{\a^\text{NS}} \big( a_0 - a_{1/2} q^{1/2} + a_{1} q^1 - a_{3/2}q^{3/2} + \cdots \big), 
  \\
  \chi^\text{R}(q) & = q^{\a^\text{R}} \big( b_0 + b_{1} q^{1} + b_{2} q^2 + b_{3}q^{3} + \cdots \big),
  \nonumber
\end{align}
where the exponents are determined by the central charge $c$  of the theory and conformal weights $h$ of primaries,
\begin{align}\label{defalpha}
    \a^\text{NS}  = h^\text{NS} - \frac{c}{24}, 
    \qquad  
    \a^\text{R}  = h^\text{R} - \frac{c}{24}.
\end{align}
Henceforth we will focus our attention to fermionic RCFTs where half-integer spin descendants 
exist in the NS sector, i.e., the Fourier coefficients $a_n$ with a half-integer $n$  
are in general non-zeroes, with some attention to the implications of the R sector.  In other words, a fermionic RCFT of our interest has an extended chiral 
algebra that contains conserved currents of half-integer spin.

The invariance of $Z_\text{NS}$, $Z_{\widetilde{\text{NS}}}$, and $Z_\text{R}$ 
under $\Gamma_\theta$, $\Gamma^0(2)$, and $\Gamma_0(2)$ implies 
that the conformal characters $\chi^\text{NS}(q) $, $\chi^{\widetilde{\text{NS}}}(q)$, 
and $\chi^\text{R}(q)$ transform as vector-valued modular forms 
for corresponding modular symmetry groups, respectively. 
The essential features of conformal characters are presented below:
\begin{itemize}
    
    \item {\bf Integrality}: All Fourier coefficients of conformal characters have to be integer-valued. In addition the modular pairing matrix $M_{ij}$ has the integral entries.  Otherwise, we have to lose the     Hamiltonian interpretation of the partition functions
    \eqref{Hamiltonian01}.
    
    \item {\bf Positivity}: In particular, the Fourier coefficients of $\chi^\text{NS}$(q) and $\chi^\text{R}$(q) are further required to be non-negative. In addition the entries  of the modular pairing matrix $M_{ij}$ should be positive.  
    
    \item {\bf Unique vacuum}: The first Fourier coefficient of the NS vacuum character has to be unit. In addition $M_{00}=1$.
    This is because, by definition, any CFT has a unique vacuum that corresponds 
    to the identity operator. 
    
    \item {\bf Weak holomorphicity}: Conformal characters in each sector 
    are holomorphic in $\tau$ inside the fundamental domain for 
    the corresponding level-two congruence subgroup of $\mathrm{SL}(2,\mathbb{Z})$. 
    
    \item {\bf Unitarity} ({\it optional}): For the unitary CFTs, the  central charge $c$ as well as
    the conformal weights $h$ should be 
    non-negative for a unitary CFT. It implies that, in the limit $\tau\to i \infty$, the 
    conformal character over the identity is the dominant one 
    in the NS sector. 
    
\end{itemize}

In order to explore the space of FRCFTs, one can naturally
study the whole space of vector-valued modular forms with the above  properties. Let us first   discuss a systematic and practical approach 
to construct such vector-valued modular forms by studying the MLDEs satisfied by these  characters\cite{Anderson:1987ge,Eguchi:1987qd,Gaberdiel:2008pr}.

\subsection{Modular linear differential equations}\label{sec:MLDE}

In this subsection, we review the holomorphic modular bootstrap method for FRCFTs. As a prerequisite, one first needs to understand how it works for bosonic cases. 

We rely on the fact that the conformal characters for 
a given bosonic RCFT are solutions to an MLDE.
The role of derivative in such an MLDE is played by the so-called Serre derivative, which maps a modular form of weight $k$ to a new modular form of weight $k+2$,
\be
D_k = \frac{1}{2\pi i}\frac{d}{d\tau} - \frac{k}{12}E_2(\tau)\,,
\ee
where $E_2(\tau)$ is the quasi-modular Eisenstein series of weight $2$. 
We define the order $n$ modular derivative as
\be
{\cal D}^{n} := D_{2n-2} \circ D_{2n-4} \cdots D_0\, , 
\ee
acting on weight zero modular forms. We choose the convention ${\cal D}^0=1$.

Let us consider an arbitrary  function $f$ made of a linear combination of $d$ characters $\{\chi_0,...,\chi_{d-1}\}$, which constitute a $d$-dimensional vector-valued modular form of weight zero. The following determinant then obviously vanishes,
\begin{align}
  {\rm det} \left( \begin{array}{cccc}
   f& \chi_0 & \cdots &\chi_{d-1}  \\
  {\cal D}^1 f &  {\cal D} \chi_0 & \cdots & {\cal D}^1 \chi_{d-1} \\
    \vdots & \vdots & & \vdots \\   {\cal D}^d f & {\cal D}^d \chi_0 & \cdots & {\cal D}^d \chi_{d-1} \\
\end{array}\right) = 0.
\end{align}
Dividing by the Wronskian $W(\tau)$ of $d$ characters $\chi_i(\tau)$, one can obtain the desired MLDE from the above determinant, 
\be\label{eq:MLDE}
  \left[ {\cal D}^d  + \sum_{k=0}^{d-1} \phi_k(\tau) {\cal D}^k \right] f(\tau) =0,
\ee
where $\phi_k(\tau)$ are modular forms of weight $(2d-2k)$ for $\mathrm{SL}(2,\mathbb{Z})$.
Since the conformal characters $\chi_i(\tau)$ are weakly holomorphic, 
$\phi_k(\tau)$ are allowed to have poles on the upper-half plane only at the zeros of the Wronskian $W(\tau)$. 

Since the order of poles is important to restrict the possible form of $\phi_k$, one introduces the notion of index $\tilde{l}$ as six times the sum of order of zeros for $W(\tau)$ in the fundamental domain. Notice that the order of zeros  
becomes $1/3$ at an orbifold point $\tau = e^{2\pi i/3}$ while $1/2$ at  
another orbifold point $\tau = i$. Furthermore, from the valence formula of $W(\tau)$, we are able to relate the parameter $\tilde{l}$ to 
the central charge $c$, conformal weights $h_i$, and the number of primaries $d$: 
\be\label{eq:lpole}
\frac{\tilde{l}}{6} = \frac{d(d-1)}{12} - \sum_{i = 0}^{d-1} \alpha_i,
\ee
where $\alpha_i = h_i - c/24$ is the leading exponent of each character in the $q$-expansion.

The core idea of holomorphic modular bootstrap is to turn the above logic around. We instead start from a discrete set of data $\{\tilde{l},d\}$ and classify all possible character-like solutions satisfying the aforementioned 
physical constraints: (i) integrality, (ii) positivity, (iii) existence of the unique vacuum, and (iv) weak holomorphicity. As a special case, the case $\tilde{l} = 0$ is dubbed the holomorphic or monic MLDE, and their complete classification for the smallest nontrivial value $d = 2$ marks the first triumph of the holomorphic modular bootstrap \cite{Mathur:1988na}.

The MLDE method was extended to classify the fermionic RCFTs recently in \cite{Bae:2020xzl,Bae:2021mej}. One can argue that 
conformal characters for each boundary condition solve 
a linear differential equation invariant under the corresponding 
level-two congruence subgroup of $\mathrm{SL}(2,\mathbb{Z})$. More precisely, 
the FMLDEs for NS, $\widetilde{\mathrm{NS}}$, and R characters 
become
\begin{gather}\label{fMLDEeq}
    \left[ {\cal D}^d  + \sum_{k=0}^{d-1} \phi_k^\text{NS} (\tau) {\cal D}^k \right] 
    f^\text{NS}(\tau)  = 0, 
    \nonumber \\ 
    \left[ {\cal D}^d  + \sum_{k=0}^{d-1} \phi_k^{\widetilde{\text{NS}}} (\tau) {\cal D}^k \right] 
    f^{\widetilde{\text{NS}}}(\tau)  = 0,
     \\ 
    \left[ {\cal D}^d  + \sum_{k=0}^{d-1} \phi_k^\text{R} (\tau) {\cal D}^k \right] 
    f^\text{R}(\tau)  = 0. \nonumber
\end{gather}
The coefficients $\phi_k^\text{NS}(\tau)$,  $\phi_k^{\widetilde{\text{NS}}}(\tau)$, and  $\phi_k^\text{R}(\tau)$ are the weight $(2d-2k)$ modular forms of the respective level-two congruence subgroups.

Analogous to \eqref{eq:lpole}, one can obtain the valence 
formula for fermionic theories by performing the contour integral of 
Wronskians from conformal characters for each boundary condition. Here the 
contour is chosen as the boundary of fundamental domain of 
the corresponding modular symmetry group.  
To be more concrete, let us consider for instance the NS sector. Its fundamental domain can be chosen as the region $\{\tau \in \mathbb{H}\big|\ |\tau| \geq 1 \ \textrm{and} \ |\textrm{Re}(\tau)| \leq 1\}$. 
There is one orbifold point after gluing the boundary, i.e., $\tau = i$ with cone angle $\pi$. 
The fundamental domain of $\Gamma_{\theta}$ has two cusps $\tau=+i\infty$ and $\tau=\pm 1$. 
Performing a contour integral along the boundary of the region, we obtain the valence formula \cite{Bae:2020xzl}
\begin{equation}\label{eq:valence}
       \frac{\ell}{2}= \frac{d(d-1)}{4} - 2 \sum_{j=0}^{d-1} \alpha_j^{\rm NS} - \sum_{j=0}^{d-1} \alpha_j^{\rm R},
\end{equation}
where $\alpha^\text{NS}$ and $\alpha^\text{R}$ are given by \eqref{defalpha}. 
Here $d$ denotes the number of characters in the NS sector and $\ell$ is the Wronskian 
index, the analog of $\tilde{l}$, counting the zeros for $W_\text{NS}$ twice except zero at $\tau = i$ only once in the fundamental domain. One interesting feature 
of \eqref{eq:valence} is that, even if we only focus on the NS sector, 
we still need information from the R sector, which arises from the cusp  $\tau=\pm 1$.

In general, one can express the coefficient functions $\phi_k(\tau)$ in \eqref{fMLDEeq}
as rational functions of Jacobi theta functions,
\begin{equation}
\vartheta_2(\tau) =  \sum_{n = -\infty}^{\infty}  q^{(n+1/2)^2/2}, \quad \vartheta_3(\tau) =  \sum_{n = -\infty}^{\infty}  q^{n^2/2}, \quad \vartheta_4(\tau) =  \sum_{n = -\infty}^{\infty} (-1)^n q^{n^2/2}.
\end{equation}
This is because $\phi_k(\tau)$ is defined as ratio of 
two holomorphic modular forms, and 
the space of entire modular forms of a given weight 
is generated by combinations of Jacobi theta functions
\footnote{We correct here a typo in equation (3.12) of \cite{Bae:2020xzl}.}: 
for non-negative integers $(r,s)$, 
\begin{align}\label{holmod}
    \mathcal{M}_{2k}(\Gamma_\theta) &=\left \langle \ (-\vartheta_2^4)^r\vartheta_4^{4s} +  (-\vartheta_2^4)^s\vartheta_4^{4r}, \ r\leq s, \ r+s=k \ \right\rangle , \nonumber\\
    \mathcal{M}_{2k}(\Gamma^0(2)) &=\left \langle \ \vartheta_2^{4r}\vartheta_3^{4s} +   \vartheta_2^{4s}\vartheta_3^{4r} ,\ r\leq s,  \ r+s=k \ \right \rangle, 
    \\
    \mathcal{M}_{2k}(\Gamma_0(2)  ) &= \left \langle \  \vartheta_3^{4r}\vartheta_4^{4s} +  \vartheta_3^{4s}\vartheta_4^{4r} , \ r\leq s, \ r+s=k \ \right \rangle . \nonumber
\end{align}
Here $\CM_{2k}(G)$ denotes the space of holomorphic modular forms 
of weight $2k$ for modular symmetry group $G$. Moreover, based on 
the transformation rules below, 
\begin{equation}
\begin{aligned}
    T:&\ (\vartheta_2^4, \vartheta_3^4, \vartheta_4^4)\ \longrightarrow \ (-\vartheta_2^4, \vartheta_4^4, \vartheta_3^4)\\
    S:&\ (\vartheta_2^4, \vartheta_3^4, \vartheta_4^4)\ \longrightarrow \ (-\tau^2\vartheta_4^4, -\tau^2\vartheta_3^4, -\tau^2\vartheta_2^4).\\
\end{aligned}
\end{equation}
we can easily read off how the coefficient functions for different 
boundary conditions are related.   

Clearly not all FMLDE solutions can be identified as conformal 
characters of putative CFTs.  The  Fourier coefficients of some solution can be rational with
 a growing denominator, and so cannot be made integral by  multiplication of a large integer.   We will discuss how the integrality 
of conformal characters can be reflected in the
representation theory of modular symmetry groups below.

\subsection{Integrality conjecture}\label{Sec:Integrality}

The integrality is the primary criterion for the solutions to 
an MLDE to be considered as physical characters. It was first 
observed in \cite{Mathur:1989pk} that the conformal characters solve an MLDE with 
finite monodromy group. Soon afterwards the integrality 
conjecture was proposed to explain the above observation. 
The conjecture states that if a vvmf transforming 
under $\mathrm{SL}(2,\mathbb{Z})$ has integer Fourier coefficients, then 
there must exist a principal congruence subgroup $\Gamma(N)$
under which each of its components becomes singlet. Here $\Gamma(N)$ is defined as 
\be
    \Gamma(N) := \Bigg\{\begin{pmatrix}
        \a& \b\\
        \g & \d
    \end{pmatrix} \in \mathrm{SL}(2,\mathbb{Z}), \quad  
    \begin{pmatrix}
         \a& \b\\ \g& \d
    \end{pmatrix} \equiv \begin{pmatrix}
        1&0\\
        0&1
    \end{pmatrix} \text{mod}\ N\Bigg\}
\ee
for some $N\in \mathbb{N}$. In other words, any $d$-dimensional vvmf with integer Fourier coefficients
has to transform in a $d$-dimensional representation of $\mathrm{SL}(2,\mathbb{Z})/\Gamma(N)=\mathrm{SL}(2,\mathbb{Z}_N)$ . 
This conjecture was recently proved in \cite{calegari2021unbounded}, and 
when the vvmf arises from a bosonic RCFT
it goes under the name of the congruence property proven earlier in \cite{Ng:2012ty}.

The irreducible representations of $\mathrm{SL}(2,\mathbb{Z}_N)$ were all classified in \cite{Nobs:1976f,Nobs:1976s,Eholzer:1994th},
and can be easily accessed from computer software such as \textbf{GAP} \cite{GAP4}.
Based on the above fact, the integrality `theorem' allows us to tightly constrain 
possible values of 
the central charge and conformal weights of vvmfs with integer Fourier coefficients for a given rank $d$. 
This method was studied in \cite{Kaidi:2021ent} to classify the space of bosonic RCFTs, 
which we review briefly below. 

Suppose $ N = \prod_i p_i^{\lambda_i} $ is factorized in terms of prime numbers $p_i$. 
We then have a finite group decomposition
\be\label{eq:SL2ndecom}
\mathrm{SL}(2,\mathbb{Z}_N) = \prod_i \mathrm{SL}(2,\mathbb{Z}_{p_i^{\lambda_i}})\,.
\ee
Since the representation of $\mathrm{SL}(2,\mathbb{Z}_N)$ also factorizes accordingly,
it is sufficient to determine a finite list of irreducible representations of 
$\mathrm{SL}(2,\mathbb{Z}_{p^\lambda})$ for any prime number $p$ and $\lambda\geq1$, out of which 
we can build up irreducible representations for each positive integer $N$.

We should stress here that the irreducible representations of 
$\mathrm{SL}(2,\mathbb{Z}_{p^\l})$ of our interest are those which 
do not arise from irreducible representations of $\mathrm{SL}(2,\mathbb{Z}_{p^{\l-1}})$ via the natural 
projection map. On the other hand, for reducible representations the modular S-matrix can be block-diagonalized, 
and there must exist two conformal weights which differs by 
an integer.\footnote{In the context of MTC this is known as the $\mathfrak{t}$-spectrum criteria \cite{bruillard2016classification}. Some constraints to construct reducible modular data from irreducible ones can be found in \cite{ng2022reconstruction}.} Therefore, we restrict ourselves to 
non-degenerate bosonic RCFTs where any difference between two conformal 
weights is not an integer. 

One important remark on the admissible representations 
of $\mathrm{SL}(2,\mathbb{Z}_N)$ is in order. 
Precisely speaking, the conformal characters transform in a representation 
of $\mathrm{PSL}(2,\mathbb{Z}_N)$ rather than $\mathrm{SL}(2,\mathbb{Z}_N)$. 
It implies that one has to search for 
irreducible representations of $\mathrm{SL}(2,\mathbb{Z}_N)$ where 
the minus of the identity acts trivially. 

As an illustration, let us consider the simplest 
but non-trivial example, $d=2$. From the representation theory 
of $\mathrm{SL}(2,\mathbb{Z}_N)$, one can show that the list of 
possible values of $N$ which allow a two-dimensional representation 
fulfilling the above requirements is finite and is given as
\begin{align}
    N \in \{2,6,8,12,20,24,60\}\,.    
\end{align}
In fact, one can argue that there are  only finitely many 
values of $N$ for which any irreducible representation in   $d$-dimension can exist. Intuitively, it can be explained 
by the fact that the irreducible 
representations of  $\mathrm{SL}(2,\mathbb{Z}_{p^{\lambda}})$
for a large $\lambda$ with sufficiently small dimensions 
all arise from pull-backs of those of  $\mathrm{SL}(2,\mathbb{Z}_{p^{\lambda_0}})$
with $\lambda_0 < \lambda$. 

In \cite{Kaidi:2021ent}, the authors carried out this procedure up to $d = 5$, 
and finally able to bootstrap candidate characters for RCFTs with Wronskian index $l < 6$. 
Motivated by their success, we would like to generalize this method to FRCFTs.

The first step is to extend the integrality theorem to the theory with fermions. 
\begin{conj}
If a $d$-dimensional vvmf is covariant under the modular group $\Gamma_{\theta}$ and has integral Fourier coefficients, then all $d$ components  are  modular functions for a fixed principal congruence subgroup $\Gamma(N) < \Gamma_{\theta}$ with some positive even integer $N$, taken to be the smallest possible.
\end{conj}
\noindent We remark that in the context of super-MTC, assuming the existence of a \textit{minimal modular extension} (which was shown to be true in \cite{johnson2021minimal}), this was proved by \cite{bonderson2018congruence}. However, the general statement in terms of vvmf seems to be still open. Also notice that $N$ here must be even, since otherwise $\Gamma(N)$ cannot be a subgroup of $\Gamma_{\theta}$.
Then the isomorphism \eqref{eq:SL2ndecom} can be generalized as follows, 
\begin{thm}\label{thm:decompose}
Assume $N = 2^k Q$ with $k > 0$ and $gcd(2,Q) = 1$. We have the following decomposition, 
\begin{equation}\label{eq:decomp}
\Gamma_{\theta}/ \Gamma(N) \cong \Gamma_{\theta}^{(k)} \times \text{SL}(2,\mathbb{Z}_Q),    
\end{equation}
where $\Gamma_{\theta}^{(k)}$ is defined to be the finite group $\Gamma_{\theta}/\Gamma(2^k)$. 
\begin{proof}
We regard $\Gamma_{\theta}/ \Gamma(N)$ as a subgroup of $\text{SL}(2,\mathbb{Z}_N) =  \text{SL}(2,\mathbb{Z}_{2^k}) \times \text{SL}(2,\mathbb{Z}_Q)$ where, according to the Chinese reminder theorem, 
the map to each factor is given by mod $2^k$ and mod $Q$ respectively. 
We first prove that $\Gamma_{\theta}/\Gamma(N)$ after modding by $Q$ is actually $\text{SL}(2,\mathbb{Z}_Q)$. 
To see this, note that there exist two integers $(a,b)$ such that $a\cdot 2+b\cdot Q = 1$ by the Bézout theorem. 
It implies that $T^1 \equiv (T^2)^a$ mod $Q$ is contained in $\Gamma_{\theta}/\Gamma(N)$ mod $Q$.
Together with $S$ element (mod $Q$), they generate the whole $\text{SL}(2,\mathbb{Z}_Q)$. 
It is then easy to check that both sides of equation \eqref{eq:decomp} have the same order, so the kernel of the map is trivial and they are isomorphic. 
\end{proof}
\end{thm}
\noindent See \cite{bonderson2018congruence} for another proof. 
This theorem says that in order to understand the irreducible representations of $\Gamma_{\theta}/ \Gamma(N)$, the essential new ingredient 
is the group $\Gamma_{\theta}^{(k)}$. We discuss a few properties of $\Gamma_{\theta}^{(k)}$ below. 

Let us begin with a useful observation. 
The order of the finite group $\text{SL}(2,\mathbb{Z}_N)$ is given by
\begin{equation}\label{order}
    \Big|\text{SL}(2,\mathbb{Z}_N)\Big| = N^3 \prod_{p| N} (1 - \frac{1}{p^2})\,, 
\end{equation}
where the product is taken over all prime numbers dividing $N$. 
Plugging $N = 2^k$ into \eqref{order}, we have $|\text{SL}(2,\mathbb{Z}_{2^k})| = 3 \times 2^{3k-2}$. Moreover, since $\Gamma_\theta$ is an index-3 subgroup of $\text{SL}(2,\mathbb{Z})$, we learn that 
\begin{align}
     \Big|\Gamma_{\theta}^{(k)}\Big| = 2^{3k-2}. 
\end{align}
In other words, $\Gamma_{\theta}^{(k)}$ is a $2$-Sylow subgroup of $\text{SL}(2,\mathbb{Z}_{2^k})$. 
It is well-known that, for a given finite group $G$, 
the total number of $p$-Sylow groups of $G$ is given by the 
index of the normalizer of any $p$-Sylow subgroup.  Moreover, 
all $p$-Sylow subgroups are conjugate to each other by certain elements of $G$. 
In our case where $G=\mathrm{SL}(2,\mathbb{Z}_{2^k})$ and $p=2$, we have three different 
$2$-Sylow subgroups that are simply the quotient of $\Gamma_{\theta}, \Gamma^0(2)$ and $\Gamma_0(2)$ by $\Gamma(2^k)$. 
One can also argue that three $2$-Sylow subgroups are conjugate to each other 
by the elements $T$ and $S$ of $\mathrm{SL}(2,\mathbb{Z}_{2^k})$, 
\begin{equation}
\Gamma_\theta^{(k)}\ \stackrel{T}{\Longleftrightarrow}\ \Gamma^0(2)/\Gamma(2^k) \stackrel{S}{\Longleftrightarrow}\ \Gamma_0(2)/\Gamma(2^k),
\end{equation}
which exactly agree with the maps between three spin structures NS, $\widetilde{\text{NS}}$, and R. 

In practice, one can identify any of them in the software \textbf{GAP} \cite{GAP4} by the command ``SylowSubgroup'' and ``ConjugateGroup''. Then its irreducible representations are easily accessed from the command ``CharacterTable''. We demonstrate it in a few examples with small $k$ below. 

Let us consider the simplest example $\Gamma_{\theta}^{(1)}$. 
Since $\Gamma_{\theta}^{(1)}$ is of order two, it is isomorphic to 
the cyclic group $\mathbb{Z}_2$. It is known that there are
two irreducible representations of dimension one. One of them is 
the trivial representation and the other is the sign representation. 
The character table of $\Gamma_{\theta}^{(1)}$ is simple, 
\[\begin{array}{lrr}
& 1a&2a\\\hline
ch_1&1 & 1 \\
ch_2&1 & -1
\end{array}\]
where each row is for an irreducible representation while 
each column for a conjugacy class. We adopt the notation 
for each conjugacy class from \textbf{GAP} where 
the prefactor shows the order of its elements. 
Since the two elements $(-I)$ and $T^2$ are equal to $I$ modulo two, 
they are in conjugacy class $1a$. Therefore, one can conclude 
that $\Gamma_{\theta}^{(1)}$ has two one-dimensional irreducible representations
where $(-I)$ acts trivially. 

Next, we move onto the first non-trivial example $\Gamma_{\theta}^{(2)}$. It 
is a finite group of order $16$. Using \textbf{GAP}, one learns that 
there are ten different conjugacy classes whose character table is 
presented below, 
\[
\begin{array}{lrrrrrrrrrr}\label{chtable2}

&1a&4a&4b&4c&4d&2a&2b&2c&2d&2e\\
\hline
ch_1&1&1&1&1&1&1&1&1&1&1\\
ch_2&1&-1&1&-1&1&-1&1&1&-1&1\\
ch_3&1&-1&-1&-1&-1&1&1&1&1&1\\
ch_4&1&1&-1&1&-1&-1&1&1&-1&1\\
ch_5&1&i&-i&-i&i&1&1&-1&-1&-1\\
ch_6&1&-i&-i&i&i&-1&1&-1&1&-1\\
ch_7&1&-i&i&i&-i&1&1&-1&-1&-1\\
ch_8&1&i&i&-i&-i&-1&1&-1&1&-1\\
ch_9&2&0&0&0&0&0&-2&2&0&-2\\
ch_{10}&2&0&0&0&0&0&-2&-2&0&2\\
\end{array}
\]
where $i = \sqrt{-1}$.
Unlike the previous example, 
we should identify which representation comes from $\Gamma_\th^{(1)}$ via the natural projection map $\Gamma_\th^{(2)}\to 
\Gamma_\th^{(1)}$. Note that the conjugacy 
classes $4a$, $4b$, $4c$, and $4d$ of $\Gamma_\th^{(2)}$ 
descend to the $2a$ class of $\Gamma_\th^{(1)}$ while 
the others to the $1a$ class. It implies that two representations 
$ch_1$ and $ch_3$ are simply the pull-back 
of the trivial and sign representation of $\Gamma_\th^{(1)}$. 
They should be disregarded as ``proper'' irreducible 
representations of $\Gamma_\th^{(2)}$ according to our assumption. 
The group $\Gamma_\th^{(2)}$ thus have $8$ ``proper'' irreducible representations in total,
two of which are two-dimensional and others are one-dimensional.
Since $(-I)$ is in the conjugacy class $2c$, one can also 
read off its action on each representation from the character table. 
For instance, $(-I)$ acts trivially on the representation $ch_9$ while
non-trivially on the representation $ch_{10}$.

In a similar fashion, after deleting those coming from smaller $k$, one proceeds to find 28  ``proper'' irreducible representations for $\Gamma_{\theta}^{(3)}$, 78 for $\Gamma_{\theta}^{(4)}$, and so on.

There are finitely many possible values of $N$ so that 
$\Gamma_\th/\Gamma(N)$ allows a $d$-dimensional irreducible 
representation where $(-I)$ acts trivially. 
According to the theorem \ref{thm:decompose}, 
a irreducible representation of $\Gamma_\th/\Gamma(N)$ 
can be described as a tensor product 
of a irreducible representation of $\Gamma_\th^{(k)}$
and a irreducible representation of $\mathrm{SL}(2,\mathbb{Z}_Q)$. 
It was shown in \cite{Kaidi:2021ent} that 
the sets of odd numbers $Q$ such that 
$\mathrm{SL}(2,\mathbb{Z}_Q)$ has a $d$-dimensional 
irreducible representation, denoted by $\mathfrak{Den}_o(d)$, are 
\begin{equation}\label{denodd}
    \begin{aligned}
    \mathfrak{Den}_o(1) &= \{1^+, 3^+\},\\
    \mathfrak{Den}_o(2) &= \{3^-, 5^- , 15^-\},\\
    \mathfrak{Den}_o(3) &= \{3^+, 5^+, 7^+, 15^+, 21^+\},\\
    \mathfrak{Den}_o(4) &= \{5^\pm, 7^-, 9^\pm, 15^\pm, {21}^-\},
    \end{aligned}
\end{equation}
where the superscript $+$ ($-$) indicates the trivial (non-trivial) action of $(-I)$ 
in a given representation. 
We also define $\mathfrak{Den}_e(d)$ as a set of $2^k$ 
such that $\Gamma_\theta^{(k)}$ has a $d$-dimensional irreducible representation 
that turns reducible under the map $\Gamma_\theta^{(k)} \to \Gamma_\theta^{(k-1)}$. 
We utilize the program \textbf{GAP} to verify that 
\begin{equation}\label{deneven}
  \begin{aligned}
  \mathfrak{Den}_e(1) &= \{2^+, 4^\pm, 8^\pm, 16^\pm\},\\
  \mathfrak{Den}_e(2) &= \{4^\pm, 8^\pm, 16^\pm, 32^\pm\},\\
  \mathfrak{Den}_e(3) &= \phi,\\
  \mathfrak{Den}_e(4) &= \{8^\pm, 16^\pm, 32^\pm, 64^\pm\}.
  \end{aligned}  
\end{equation}
In general, we claim that $\mathfrak{Den}_e$ is non-empty if and only if $m$ has only prime factor 2, and we give a proof in Appendix \ref{App:Group}. Based on \eqref{denodd} and \eqref{deneven}, it is straightforward 
to compute the possible values of $N$ for which $\Gamma_\th/\Gamma(N)$ has irreducible representations 
where $(-I)$ has trivial action. We summarize the results in the Table \ref{ListofN}. 
\begin{table}[t!]
\centering
\def\arraystretch{1.3}
%\scalebox{1.0}{
\begin{tabular}{l | l | l}
\hline 
 $d$ & $N$ & $\text{Number of irrep.}$\\
\hline 
$1$ & $\big\{2,4,6,8,12,16,24,48 \big\}$ & $48$ irreps.  \\ [1.5mm] 
$2$ & $\big\{4,8,12,16,20,24,32,40,48,60,80,96,120,240 \big\}$ & $300$ irreps. \\ [1.5mm] 
\multirow{2}{*}{$3$} & $\big\{ 6,10,12,14,20,24,28,30,40,42,48,56,60,80,84,$ & \multirow{2}{*}{$208$ irreps.} \\ [1mm]
& $112,120,168,240,336\big\}$ & \\ [1.5mm]
\multirow{2}{*}{$4$} & $\big\{ 8, 10, 12, 16, 18, 20, 24, 28, 30, 32, 36, 40, 48, 56, 60,$ & \multirow{2}{*}{$1206$ irreps.} \\ [1.5mm]
& $64, 72, 80, 84, 96, 112, 120, 144, 160, 168, 192, 240, 336, 480\big\} $ & \\ [1.5mm]
\hline
\end{tabular}
%}
\caption{\label{ListofN} Complete lists of allowed values of $N$ for 
FRCFTs as well as the number of possible irreducible representations when $d=1,2,3,4$  }
\end{table}

\subsection{Candidate conformal weights}
We make use of the integrality conjecture to classify 
the possible values of $N$ such that putative 
conformal characters $\chi_i^\text{NS}$ transform in an $d$-dimensional irreducible 
representation $\r$ of $\Gamma_\th/\Gamma(N)$. 
Since the element $T^2$ of $\Gamma_\th/\Gamma(N)$ 
has to satisfy the relation $(T^2)^{N/2}=T^N=1$, 
we see that
\begin{align}
    \chi_i^\text{NS} (\tau+N) = e^{2\pi i (N \alpha_i^\text{NS}) } \chi_i^\text{NS} (\tau) 
    = \chi_i^\text{NS} (\tau), 
\end{align}
where the exponents are defined in \eqref{exponents}, $\alpha_i^\text{NS} = h_i^\text{NS} - c/24$. 
The value of $N$ thus determines the denominator of the exponents. 
In fact, the representation theory of $\Gamma_\th/\Gamma(N)$ 
can further constrain the space of allowed central charges and conformal weights.
This is because each exponent $\alpha_i^\text{NS}$ gives rise to an 
eigenvalue of $\rho(T^2)$,
\begin{align}
    \rho(T^2) ~\dot{=}~ 
        \begin{pmatrix}
        e^{4\pi i  \alpha^\text{NS}_0} & & &  
        \\ 
        & e^{4\pi i  \alpha^\text{NS}_1 } & &  
        \\
        & & \ddots & \\
        & & &  e^{4\pi i  \alpha^\text{NS}_{d-1} } 
    \end{pmatrix}. 
\end{align}
We shall argue that those eigenvalues can be read off 
from the character table.

Suppose that the conformal characters 
are in a representation $\r$ of $\Gamma_\th/\Gamma(N)$. The eigenvalues of $\rho(T^2)$ are 
then all of the form $\exp{(\frac{4\pi i m }{N})}$ for some integer $m$ between $1$ and $N/2$,
due to the relation $\rho(T^2)^{N/2} = 1$. 
To determine those eigenvalues, it is essential to identify 
the conjugacy classes of $(T^2)^l$ for $l=1,2,..N/2$. As a result, we can 
compute the characters of $(T^2)^l$ from the character table. Each characters 
can be also expressed in terms of eigenvalues as,
\begin{align}\label{determineeigenvalues}
    ch_\r\big[  (T^2)^l \big] \equiv \text{Tr}_\rho\big[ (T^2)^l \big]  = \sum_{m=1}^{N/2} 
    \nu^{(\rho)}_m e^{ \frac{4\pi i m l}{N} }, 
\end{align}
where $\nu^{(\rho)}_m$ counts the degeneracy of  $\exp{(\frac{4\pi i m}{N})}$ in $\rho(T^2)$. 
Invoking the inverse discrete Fourier transform, it is straightforward to solve \eqref{determineeigenvalues},
\begin{equation}\label{eq:inverseFourier} 
    \nu^{(\rho)}_m = \frac{2}{N} \sum_{j = 1}^{N/2}  ch_\r\big[  (T^2)^j \big] \cdot e^{ \frac{-4\pi i m j}{N} },
\end{equation}
from which one can reconstruct  the eigenvalues of $\rho(T^2)$. 

We demonstrate the above procedure for the example $\Gamma_\theta^{(2)}$, i.e., $N=4$. 
Using \textbf{GAP}, we learn that the element $T^2$ lies in the class $2a$. 
When the conformal characters $\chi_i$ ($i=0,1$) are in the two-dimensional 
representation with trivial $(-I)$ action, the characters of $T^2$ and $T^4=I$ read
\begin{align}
 ch_9\big[I \big] = 2 =  \sum_{m=1}^2 \nu_m^{(9)}, \quad 
 ch_9\big[T^2\big] =    0 = \sum_{m=1}^2  (-1)^m \nu_m^{(9)} .  \nonumber 
\end{align}
The matrix $\rho(T^2)$ therefore becomes 
\begin{align}
    \rho(T^2) = 
    \begin{pmatrix}
        1 & 0 \\ 0 & -1
    \end{pmatrix},
\end{align}
and thus $2(\alpha^\text{NS}_0, \alpha^\text{NS}_1) = ( 0 , 1/2) $ modulo integers. 

Remarkably, the character table of $\Gamma_\th/\Gamma(N)$
is not ignorant of information from the R boundary condition. To be more concrete,   
the characters of group elements $TS.T^l.(TS)^{-1}$ for $l=1,2,..,N$ determine
the conformal weights $h^\text{R}_i$ or equivalently the exponents $\a_i^\text{R}$. 
To see this, we first note that a representation of $\Gamma_\th/\Gamma(N)$ can induce representations of $\text{SL}(2,\mathbb{Z}_N)$. They all give rise to the same result so we just pick one of them and continue to call it $\rho$. Then we can regard the above elements in $\Gamma_\th/\Gamma(N)$ as genuine product of elements in $\text{SL}(2,\mathbb{Z}_N)$, and we have
\begin{align}
    \text{Tr}_{\rho|_\text{NS}} \Big[ TS.T^l.(TS)^{-1} \Big] = 
    \text{Tr}_{\rho|_\text{R}} \Big[T^l \Big].
\end{align}
This is because $(TS)^{-1}$ maps any vector in $\rho|_\text{NS}$ 
to a vector in $\rho|_\text{R}$, i.e., $| i \rangle_\text{NS} = \sum_{j} (\CT\CS)_{ij} |j\rangle_\text{R}$. 
Once we identify the conjugacy classes of $\Gamma_\th/\Gamma(N)$ where $TS.T^l.(TS)^{-1}$ are placed, 
the eigenvalues of $\rho|_\text{R}(T)$ as well as the exponents $\alpha^\text{R}_i$ modulo integers 
\begin{align}
    \rho|_\text{R}(T) ~\dot{=}~ 
        \begin{pmatrix}
        e^{2\pi i  \alpha^\text{R}_0} & & &  
        \\ 
        & e^{2\pi i  \alpha^\text{R}_1 } & &  
        \\
        & & \ddots & \\
        & & &  e^{2\pi i  \alpha^\text{R}_{d-1} } 
    \end{pmatrix} 
\end{align}
can be thus worked out in the similar fashion.

Again, let us consider the case $\Gamma_\theta^{(2)}$ for illustration.
For the two-dimensional representation that is singlet under $(-I)$, 
the eigenvalues of $\rho(TS.T.(TS)^{-1})$ can be expressed as 
$\exp{(\frac{2 \pi i m}{4})}$ for some integer $m$ between $1$ and $4$.
Let us denote by $\n_m^{(9)}$ the number of times the eigenvalue $\exp{(\frac{  i\pi  m}{2})}$ 
occurs. We utilize the program \textbf{GAP} to see that 
the conjugacy classes of $TS.T^l.(TS)^{-1}$ for $l=1,2,3,4$ 
are $4a$, $2e$, $4c$, and $1a$ respectively. 
From the character table 
of $\Gamma_\theta^{(2)}$, one can read off the corresponding 
characters, 
\begin{gather}
 ch_9[I] = 2 = \sum_{m=1}^4 \n_m^{(9)}, \quad  
 ch_9[TS.T.(TS)^{-1}] = 0 = \sum_{m=1}^4 (i)^{m} \n_m^{(9)}, 
 \nonumber \\ 
 ch_9[TS.T^2.(TS)^{-1}] = -2 = \sum_{m=1}^4 (-1)^{m} \n_m^{(9)}, \quad
  ch_9[TS.T^3.(TS)^{-1}] = 0 = \sum_{m=1}^4 (-i)^{m} \n_m^{(9)}.
 \nonumber
\end{gather}
It implies that the eigenvalues of $\rho_9(TS.T.(TS)^{-1})$, namely 
those of $\rho|_\text{R}(T)$, are $\pm i$. In other words, 
$(\alpha_0^\text{R},\alpha_1^\text{R}) = (1/4,3/4)$ modulo integers. 

All we need is to carry out the procedure reading off exponent pairs $(2\alpha^\text{NS},\alpha^\text{R})$ 
for all possible representations in Table \ref{ListofN}. Let us start with the case of rank one, $d=1$. 
The results already appear intriguing. We find that the $48$ representations for $d=1$ 
give rise to two families of exponent pairs. They are summarized in Table \ref{Tab:exprankone}.
\begin{table}[t!]
    \begin{align}
     \def\arraystretch{1.5}
     \newcolumntype{?}{!{\vrule width 2pt}}
   \begin{array}{c?c}
        \hline
       \multicolumn{2}{c}{d = 1\ \text{exponents} }\\
       \hline\hline
       \text{No}. &\{2\alpha^{\text{NS}},\alpha^{\text{R} }\}\\\hline
    1 &\ \{\frac{k}{24},\frac{24-k}{24}\},\quad 0\leq k \leq 23\\\hline
    2 &\ \{\frac{k}{24},\frac{12-k}{24}\}, \quad  0\leq k \leq 23\\\hline
    \end{array}
    \end{align}
    \caption{All possible exponents mod $1$ for rank $1$ FRCFTs.}\label{Tab:exprankone}
\end{table}

One can immediately see that the exponent pairs of $k$ copies of free Majorana fermion 
are in perfect agreement with those in the first family. The torus partition functions 
of a free fermion are given by 
\begin{gather}\label{freef}
    Z_\text{NS} = \left| \sqrt{\frac{\vartheta_3(\tau)}{\eta(\tau)}} \right|^2 , \quad
    Z_{\widetilde{\text{NS}}} = \left| \sqrt{\frac{\vartheta_4(\tau)}{\eta(\tau)}} \right|^2, 
   \quad
    Z_\text{R} = \left| \sqrt{\frac{\vartheta_2(\tau)}{\eta(\tau)}} \right|^2, \quad
    Z_{\widetilde{\text{R}}}= 0,
\end{gather}
where $\theta_i(\tau)$ are Jacobi theta functions and $\eta(\tau)$ is the Dedekind eta function. The partition 
function in the $\widetilde{\text{R}}$ boundary condition vanishes because of 
fermionic zero mode. We can see that the theory of a free fermion 
is RCFT with a single character that transforms under $S$ and $T$ as follows, 
\begin{align}\label{eq:STtransform}
    & T: \chi^{\textrm{NS}}(\tau+1) = e^{-2\pi i /48} \chi^{\widetilde{\textrm{NS}}}(\tau), \quad 
    \chi^{\textrm{R}}(\tau + 1) =  e^{2\pi i /24} \chi^{\textrm{R}}(\tau),
    \nonumber \\ 
    & S: \chi^{\widetilde{\textrm{NS}}}(-1/\tau) =  \chi^{\textrm{R}}(\tau), \quad 
    \chi^{\textrm{NS}}(-1/\tau) = \chi^{\textrm{NS}}(\tau).
\end{align}
Thus we see that the eigenvalue of $\chi_{\textrm{NS}}$ under $T^2$ together with the eigenvalue of $\chi_{\textrm{R}}$ under $T$ realizes the exponent pair $\{2\alpha^\text{NS}, \alpha^\text{R}\}=\{\frac{23}{24},\frac{1}{24}\}$ in the first family.

Meanwhile we notice that there is also a second family in Table \ref{Tab:exprankone}. 
The theories in the second family share 
the same NS sector exponents $2\alpha^\text{NS}$ with those in the first.  
However the R sector exponents $\alpha^\text{R}$ differ by 1/2 pairwise. 
The difference also concerns with the action of $S$. As we can see from 
\eqref{eq:STtransform}, $S$ acts trivially on arbitrary copies of $\chi_\text{NS}$.  
On the other hand, $S$ acts as $(-1)$ for theories in the second family. 
The non-trivial action of $S$ has a physical consequence. The theory of 
a free Majorana-Weyl fermion cannot be a genuine two-dimensional CFT by its own 
because of the gravitational anomaly. However, the three-dimensional 
fermionic gravitational Chern-Simons coupling cancels the anomaly, more precisely 
the extra phases of \eqref{eq:STtransform}. It suggests that the chiral theory 
of a free Majorana-Weyl fermion can exist on the boundary of a three-dimensional 
fermionic gapped system. This is not the case for the theories in the second family
\footnote{We thank Ying-Hsuan Lin for discussion on this point.}: 
the chiral part of them cannot be realized as the boundary theories since 
$S$ acts non-trivially. Furthermore, we will confirm in the next section that 
no physical theories can realize the exponents in the second family. 

After treating the rank one theories as a warm-up, let us move on to the higher rank cases.
Repeating the above procedure, we extract all the exponent pairs from the representations in Table \ref{ListofN}.
Let us decompose the exponent pairs into families within which they are 
related by constant shifts; when two exponent pairs, $\{2\alpha^{\text{NS}},  \alpha^\text{R}\}$
and $\{2 \beta^{\text{NS}}, \beta^\text{R}\}$ are in the same family, their difference is 
\begin{align}\label{categorization}
    \{2\alpha^{\text{NS}},  \alpha^\text{R}\} - \{2 \beta^{\text{NS}}, \beta^\text{R}\}  = 
    \frac{k}{24}\{ {\bf 1}_d, - {\bf 1}_d \} \text{ mod } \mathbb{Z} , 
\end{align}
for some integer $k$ between $0$ and $23$. For this categorization, we are 
motivated by the fact that two RCFTs realizing the exponent pairs obeying \eqref{categorization}
could be potentially related by adding or subtracting $k$ copies of free fermions. 
We present all possible families for $d=2,3$ and $4$ in Appendix \ref{App:exponents}, and 
provide a representative exponent pair for each family.

\section{Classification}\label{sec:3}

So far we investigated the space of possible exponent pairs $\{2\alpha^\text{NS},  \alpha^\text{R}\}$,
or conformal weights, of vvmfs with integral Fourier coefficients, 
relying on the representation theory of $\Gamma_\th/\Gamma(N)$. 
However the integrality conjecture only determines 
the exponents pairs modulo integers. Moreover, not every vvmf 
for each choice of allowed exponent pairs can be identified 
as conformal characters satisfying the aforementioned constraints such as 
positivity, the existence of unique vacuum, and weak holomorphicity. 

The classification of FRCFTs thus needs more 
elaboration. To achieve the goal, 
we solve the FMLDEs \eqref{fMLDEeq} armed with exponent pairs in Appendix \ref{App:exponents}, 
and then analyze the solutions carefully to see 
whether physical constraints are obeyed. 
In this classification, we only consider non-degenerate theories, meaning that there exists no pair of NS sector conformal weights whose difference is a multiple of a half-integer, and no pair of R sector conformal weights whose difference is an integer. For those theories, the presentation of $\Gamma_\th/\Gamma(N)$ should be irreducible.\footnote{We expect a corresponding theorem in the super-MTC. See also footnote 2.}

To make our approach available, we restrict our attention to FMLDEs 
where the coefficient functions $\phi_k(\tau)$ are completely fixed
for a given choice of $\{2\alpha^\text{NS},  \alpha^\text{R}\}$. They are referred to as
{\it rigid} FMLDEs in the present work. 
The strategy to fix the unknown coefficients is simple. 
We first write down the FMLDE of order $d$ for $\Gamma_\theta$, and require 
the solutions to start with $\chi_j^{\text{NS}} \sim q^{\alpha_j^{\text{NS}}}(1 + \cdots)$ ($j=1,2,..,d$)
in the limit $\tau\to i \infty$. It provides us $d$ constraints among the coefficient functions. 
Then we transform the FMLDE to the one for $\Gamma_0(2)$ via a suitable modular transformation. 
Again, having solutions that begin with $\chi_j^{\text{R}} \sim q^{\alpha_j^{\text{R}}}(1 + \cdots)$ 
gives rise to another $d$ constraints. 
We need to check whether the exponents leads to independent constraints. 

When we search for unitary fermionic RCFTs,
the no-free-fermion condition can be further imposed
without loss of generality. In practice, we simply demand 
the vacuum character to have vanishing second Fourier 
coefficient, 
\begin{equation}\label{nofreefermion}
    \chi_0^{\rm NS} = q^{-c/24} (1 + 0\cdot q^{1/2} +  a_1 q^1 + \cdots).
\end{equation}
To see this, we note that  any Virasoro primary of dimension $1/2$ 
and spin $1/2$ corresponds to a free fermion. See \cite{Lee:2019uen} for a demonstration. 
The number of such primaries are counted by the 
the second Fourier coefficient of the vacuum character, specified 
by the smallest NS exponent. 

Let us finally discuss how to fully determine the exponent pairs with no integer ambiguity,
that play an essential role in our approach. We start with 
an exponent pair defined mod $\ell$ presented in Appendix \ref{App:exponents}, and 
denote it by $\{2\tilde{\a}^\text{NS},  \tilde{\a}^\text{R}\}$  for later convenience. 
We here reduce all of exponents to lie within the range $(0, 1)$, i.e., 
\begin{align}
  0 \leq 2\tilde{\a}^\text{NS}_i , \tilde{\a}^\text{R}_i   < 1 \text{ for } i=0,1,..,d-1. 
\end{align}
Then we consider all possible ways to shift each of the exponents by an integer with absolute value smaller than a fixed bound, while satisfying the valence formula \eqref{eq:valence} for a given Wronskian index $l$. In other words, 
for a set of integers $(n^\text{NS}, n^\text{R})$, we have 
\begin{equation}
    \sum_{j=0}^{d-1}\left(  2\tilde{\alpha}_j^{\text{NS}} + n^\text{NS}_j \right) 
    + \sum_{j=0}^{d-1} \left( \tilde{\alpha}_j^{\text{R}} + n^\text{R}_j \right)  
    + \frac{\ell}{2} = \frac{d(d-1)}{4}\,,
\end{equation}
where integers are bounded, $|n_j^\text{NS},n_j^\text{R}|\leq R$ for some integer $R$.  
The bound $R$ will be henceforth referred as the range. This procedure will generate a whole list of well-defined exponent sets $\{2\alpha^{\rm NS},\alpha^{\rm R}\}=\{2\tilde{\a}^\text{NS},  \tilde{\a}^\text{R}\} + \{ n^\text{NS}, n^\text{R}\}$ for a fixed range $R$. 
Then we solve the FMLDE for each of them, and expand the solutions up to roughly the order $q^{50}$. 
The next step is to impose the physical constraint listed in Section \ref{sec:MLDE}, 
and finally obtain a small list of possible candidate conformal characters.
We repeat this procedure while gradually increasing the value of $R$ until no physical 
solution can be found. 

For the case of $d=1$, we in fact do not find any physical solutions in the numerical search for 
the second family in Table \ref{Tab:exprankone}. 
This serves as a consistency check for our argument in the last section. 
Starting from rank two there are many more possibilities, which will be detailed in the following sections.

%%%%%%%%%%%%%%%%%%%%%%%%%%%%%%%%%%%%%%%%%%%%%%%%%%%%%%%%%%%%%%%%%%%%%%%%%%%%%%%%%%%%%%%%%%%%%
%%%%%%%%%%%%%%%%%%%%%%%%%%%%%%%%%%%%%%%%%%%%%%%%%%%%%%%%%%%%%%%%%%%%%%%%%%%%%%%%%%%%%%%%%%%%%

\subsection{Rank $2$ with $\ell \leq 1$} 

The second-order FMLDE for $\Gamma_\th$ \eqref{fMLDEeq} involves two 
coefficient functions $\phi_0^\text{NS}(\t)$, $\phi_1^\text{NS}(\t)$. 
When the Wronskian index $\ell \leq 1$, the coefficient functions $\phi_k(\t)$ 
are allowed to have a pole only at $\tau=i$. Since 
$(\vartheta_4^4-\vartheta_2^4)$ vanishes precisely at $\tau = i$, 
one can show from \eqref{holmod} that the coefficient functions can 
be expressed as follows,
\begin{align}
    \phi_0^\text{NS}(\t) & = \mu_3 (\vartheta_2^8 + \vartheta_4^8) + \mu_4 \vartheta_2^4 \vartheta_4^4\, , 
    \nonumber \\ 
    \phi_1^\text{NS}(\t) & = \frac{\mu_1 (\vartheta_2^8 + \vartheta_4^8) + \mu_2 \vartheta_2^4 \vartheta_4^4}{\vartheta_4^4-\vartheta_2^4}\, . 
\end{align} 
Thus, the most general FMLDE of $d=2$ with $\ell \leq 1$ for the NS sector becomes 
\begin{equation}\label{order2rigid01}
    \Bigg[\mathcal{D}^2+ \frac{\mu_1 (\vartheta_2^8 + \vartheta_4^8) + \mu_2 \vartheta_2^4 \vartheta_4^4}{\vartheta_4^4-\vartheta_2^4} \mathcal{D} + \mu_3 (\vartheta_2^8 + \vartheta_4^8) + \mu_4 \vartheta_2^4 \vartheta_4^4 \Bigg] \chi^{\text{NS}}(\tau)=0, 
\end{equation}
Performing a suitable $\text{SL}(2,\mathbb{Z})$ transformation, one can obtain 
the FMLDE for the R sector,  
\begin{equation}\label{order2rigid02}
    \Bigg[\mathcal{D}^2+ \frac{\mu_1 (\vartheta_4^8 + \vartheta_3^8) - \mu_2 \vartheta_4^4 \vartheta_3^4}{-\vartheta_3^4-\vartheta_4^4} \mathcal{D} + \mu_3 (\vartheta_4^8 + \vartheta_3^8) - \mu_4 \vartheta_4^4 \vartheta_3^4 \Bigg]\chi^{\text{R}}(\tau)=0.
\end{equation}
Since four free parameters $\mu_a$ ($a=1,2,3,4$) can be solely determined by the exponent data $\{2\a^\text{NS}_i, \a^\text{R}_i\}$ ($i=1,2$),
we see that \eqref{order2rigid01} and \eqref{order2rigid02} are rigid. 
For higher values of $\ell$, each coefficient function can have poles anywhere inside the fundamental domain, which brings in 
more undetermined parameters. Thus, the FMLDEs with $\ell>1$ cannot be rigid. 

Following the procedure outlined above, we are able to pin down $31$ solutions which can be regarded as 
physical conformal characters. The search was completed when $R=7$. 
Interestingly, even though we start with the $\ell=1$ FMLDEs, all the physically sensible solutions turn out to satisfy $\ell = 0$.

We divide the results into two groups. The first group is for unitary theories without any free fermion 
summarized in Table \ref{tab:rktwo1}. 
\begin{table}[t!]
\centering
     \def\arraystretch{1.7}
     \newcolumntype{?}{!{\vrule width 1.5pt}}
   \begin{tabular}{c|c|c}
      \hline
      \multicolumn{3}{c}{\text{Rank two, unitary},\ $\ell=0$,  $\{c, h_1^{\text{NS}}, h_0^{\text{R}}, h_1^{\text{R}}\}_{\text{type}}$ }\\
    \hline\hline
        $\left\{\frac{7}{10},\frac{1}{10},\frac{3}{80},\frac{7}{16}\right\}_B$ & 
        $\left\{11,\frac{5}{6},\frac{11}{24},\frac{9 }{8}\right\}_S$ &
        $\left\{21,\frac{5}{4},\frac{3}{2},\frac{7}{4}\right\}_B$ \\ 
    \hline
        $\left\{\frac{3}{4},\frac{1}{4},\frac{1}{32},\frac{5}{32}\right\}_S$  &
        $\left\{\frac{133}{10},\frac{9}{10},\frac{57}{80},\frac{21}{16}\right\}_B$ & 
        $\left\{\frac{85}{4},\frac{5}{4},\frac{51}{32},\frac{55}{32}\right\}_B$ \\
    \hline
        $\left\{1,\frac{1}{6},\frac{1}{24},\frac{3 }{8}\right\}_S$ & 
        $\left\{\frac{91}{5},\frac{11}{10},\frac{49}{40},\frac{13}{8}\right\}_B$ & 
        $\left\{22,\frac{4}{3},\frac{3}{2},\frac{11}{6}\right\}_B$ \\
    \hline
        $\left\{\frac{9}{4},\frac{1}{4},\frac{3}{32},\frac{15 }{32}\right\}_S$ &
        $\left\{\frac{39}{2},\frac{7}{6},\frac{65}{48},\frac{27}{16}\right\}_B$ & 
        $\left\{\frac{114}{5},\frac{7}{5},\frac{3}{2},\frac{19}{10}\right\}_B$ \\ 
    \hline
        $\left\{\frac{39}{4},\frac{3}{4},\frac{13}{32},\frac{33 }{32}\right\}_S$ & 
        $\left\{\frac{102}{5},\frac{6}{5},\frac{3}{2},\frac{17}{10}\right\}_B$ &
        \\
    \hline
    \end{tabular}
\caption{Rank 2 non-degenerate unitary theories with $\ell = 0$, which are all discovered in \cite{Bae:2020xzl}. Type: S=SUSY, B=SUSY Broken.}\label{tab:rktwo1}
\end{table}
We observe that they all have non-negative Ramond exponents, 
\begin{align}\label{SUSYUnitarityCondition}
    \alpha^\text{R}_i = h^\text{R}_i -\frac{c}{24} \geq 0 \text{ for all } i,      
\end{align}
which coincides with the unitarity bound of supersymmetry. 
It was argued in \cite{Bae:2021jkc} that any unitary FRCFT satisfying \eqref{SUSYUnitarityCondition}
is supersymmetric unless it has free fermions. When the bound is saturated, there exist 
supersymmetric Ramond vacua. Otherwise, the supersymmetry is spontaneously broken. 
We can indeed show that 
the theories in Table \ref{tab:rktwo1} are supersymmetric theories, which match all 
the rank-two non-degenerate unitary theories found in \cite{Bae:2020xzl}.  
However, we do not have a clear understanding 
why the rank-two fermionic unitary RCFTs with no free fermion should be all supersymmetric.

The results in the second group are listed in Table \ref{tab:rktwo2}.
\begin{table}[t!]
\centering
     \def\arraystretch{1.7}
     \newcolumntype{?}{!{\vrule width 1.5pt}}
   \begin{tabular}{c|c}
      \hline
      \multicolumn{2}{c}{\text{Rank two, more\,}, $ \ell=0, \{c, h_1^{\text{NS}}, h_0^{\text{R}}, h_1^{\text{R}}\}$ }\\
%       \multicolumn{2}{c}{\text{\begin{itemize}
%\color{red}
%\item  
%\end{itemize}}}\\
       \hline\hline
            $\left\{\frac{8}{5},\frac{3}{10},\frac{1}{20},\frac{1}{4}\right\}$ & 
            $\left\{33,\frac{7}{4},\frac{9}{4},3\right\}$ \\
        \hline
            %\cellcolor{gray!20}
            $\left\{2,\frac{1}{16},\frac{3}{16},\frac{11}{16}\right\}$ & 
            $\left\{\frac{341}{10},\frac{9}{5},\frac{37}{16},\frac{249}{80}\right\}$  \\
        \hline
            %\cellcolor{gray!20}
            $\left\{4,\frac{1}{8},\frac{3}{8},\frac{7}{8}\right\}$ & 
            $\left\{\frac{75}{2},\frac{7}{6},\frac{125}{48},\frac{79}{16}\right\}$ \\
        \hline
            %\cellcolor{gray!20}
            $\left\{6,\frac{3}{16},\frac{9}{16},\frac{17}{16}\right\}$ & 
            $\left\{39,\frac{5}{4},\frac{21}{8},\frac{41}{8}\right\}$ \\ 
        \hline 
            $\left\{\frac{69}{10},\frac{7}{10},\frac{1}{80},\frac{13 }{16}\right\}$ & 
            $\left\{\frac{169}{4},\frac{5}{4},\frac{99}{32},\frac{175}{32}\right\}$ \\
        \hline
            %\cellcolor{gray!20}
            $\left\{8,\frac{1}{4},\frac{3}{4},\frac{5}{4}\right\}$ &
            $\left\{\frac{87}{2},\frac{13}{6},\frac{161}{48},\frac{59}{16}\right\}$ \\
        \hline
            $\left\{\frac{151}{5},\frac{8}{5},\frac{17}{8},\frac{109}{40}\right\}$ & 
             $\left\{\frac{179}{4},\frac{9}{4},\frac{109}{32},\frac{121}{32}\right\}$ \\
        \hline
            $\left\{\frac{63}{2},\frac{5}{3},\frac{35}{16},\frac{137}{48}\right\}$ &
           $\left\{46,\frac{5}{4},\frac{29}{8},\frac{47}{8}\right\}$ \\
        \hline
            $\left\{\frac{162}{5},\frac{17}{10},\frac{11}{5},3\right\}$ & 
            $~$ \\
        \hline
\end{tabular}
\caption{Rank 2 non-degenerate theories with $\ell = 0$ that are not unitary and seem to be new. %We highlight the theories that may allow the interpretation as non-unitary theories.
}\label{tab:rktwo2}
\end{table}
They were not discovered in the previous work \cite{Bae:2020xzl}.
If the theories in Table \ref{tab:rktwo2} were unitary, they contained multiple copies of free fermion.
However, when such free fermions were decoupled, the leftover conformal characters in turn had 
negative Fourier coefficients.  
Thus, they should not be identified as
unitary theories. On the other hand, they still satisfy the constraint on the modular pairing matrix $M_{ij}$. Although it is likely that the theories of Table \ref{tab:rktwo2} 
could play no physical role, we report the results for completeness. 

Finally, we comment that all families in Table \ref{Tab:expranktwo} of Appendix \ref{App:exponents} are realized in Tables \ref{tab:rktwo1} and \ref{tab:rktwo2}.

%%%%%%%%%%%%%%%%%%%%%%%%%%%%%%%%%%%%%%%%%%%%%%%%%%%%%%%%%%%%%%%%%%%%%%%%%%%%%%%%%%%%%%%%%%%%%
%%%%%%%%%%%%%%%%%%%%%%%%%%%%%%%%%%%%%%%%%%%%%%%%%%%%%%%%%%%%%%%%%%%%%%%%%%%%%%%%%%%%%%%%%%%%%
                      
\subsection{Rank 3 with $\ell \leq 1$}

We can argue that the most general rigid FMLDE at rank three 
for the NS sector are given by 
\begin{align}\label{3rdMLDEth}
    \Bigg[ 
    \mathcal{D}^3 & + \frac{\mu_1 (\vartheta_2^8 + \vartheta_4^8) + \mu_2 \vartheta_2^4 \vartheta_4^4}{\vartheta_4^4-\vartheta_2^4} \mathcal{D}^2 +  \big(\mu_3 (\vartheta_2^8 + \vartheta_4^8) + \mu_4 \vartheta_2^4 \vartheta_4^4\big)\mathcal{D} 
    \nonumber \\
    & + \frac{\mu_5 (\vartheta_2^{16} + \vartheta_4^{16}) + \mu_6(\vartheta_4^{12}\vartheta_2^4 + \vartheta_2^{12}\vartheta_4^4) + \mu_7 \vartheta_2^8 \vartheta_4^8}{\vartheta_4^4-\vartheta_2^4} 
    \Bigg] \chi^{\text{NS}}(\tau)=0.
\end{align}
The corresponding equation for the R sector then becomes
\begin{align}
    \Bigg[
    \mathcal{D}^3 & + \frac{\mu_1 (\vartheta_4^8 + \vartheta_3^8) + \mu_2 \vartheta_4^4 \vartheta_3^4}{-\vartheta_3^4-\vartheta_4^4} \mathcal{D}^2 +  \big(\mu_3 (\vartheta_4^8 + \vartheta_3^8) - \mu_4 \vartheta_4^4 \vartheta_3^4\big)\mathcal{D} 
    \nonumber \\ 
    & + \frac{\mu_5 (\vartheta_4^{16} + \vartheta_3^{16}) -\mu_6(\vartheta_3^{12}\vartheta_4^4 + \vartheta_4^{12}\vartheta_3^4) + \mu_7 \vartheta_4^8 \vartheta_3^8}{-\vartheta_3^4-\vartheta_4^4} \Bigg] \chi^{\text{R}}(\tau)=0.
\end{align}
Note that the Wronskian index of \eqref{3rdMLDEth} is $\ell\leq1$; 
otherwise the corresponding FMLDE becomes non-rigid.  
In order to avoid having too many solutions, we only search for unitary theories in what follows. 
The seven unknown parameters $\mu_a$ ($a=1,2,..,7)$ can be specified by 
$\{2\a^\text{NS}_i, \a^\text{R}_i\}$ ($i=1,2,3$) and no-free-fermion condition 
\eqref{nofreefermion}.  

We observe that the number of unitary solutions are actually finite. 
In practice, for $\ell = 0$ cases we obtain no more solution when the range is increased from seven to eight 
while for $\ell = 1$ all the solutions are within range three. 
Altogether we obtain $43$ solutions for $\ell= 0$ and $15$ solutions for $\ell = 1$, 
listed in Tables \ref{Tab:rank3l0}
\begin{table}[t!]
\centering
     \def\arraystretch{1.7}
     \newcolumntype{?}{!{\vrule width 1.5pt}}
   \begin{tabular}{c|c|c}
      \hline
      \multicolumn{3}{c}{\text{Rank three, $\ell=0$},$\ \{c, h_1^{\text{NS}}, h_2^{\text{NS}}, h_0^{\text{R}}, h_1^{\text{R}},h_2^{\text{R}} \}_{\text{type}}$ }\\
       \hline\hline
            $\left\{\frac{11}{14},\frac{1}{14},\frac{3}{14},\frac{3}{112},\frac{5}{16},\frac{99}{112}\right\}_N$ & $\left\{\frac{77}{5},\frac{7}{10},\frac{11}{10},\frac{33}{40},\frac{49}{40},\frac{13}{8}\right\}_B$ & $\left\{\frac{187}{7},\frac{13}{14},\frac{25}{14},\frac{79}{56},\frac{17}{8},\frac{143}{56}\right\}_B$ \\
        \hline
            $\left\{1,\frac{1}{10},\frac{2}{5},\frac{1}{40},\frac{9}{40},\frac{5}{8}\right\}_N$ & $\left\{\frac{33}{2},\frac{9}{10},\frac{11}{10},\frac{77}{80},\frac{93}{80},\frac{25}{16}\right\}_B$ & $\left\{\frac{323}{10},\frac{17}{10},\frac{19}{10},\frac{31}{16},\frac{171}{80},\frac{187}{80}\right\}_B$\\
        \hline
            \cellcolor{gray!20}$\left\{\frac{7}{5},\frac{1}{10},\frac{1}{5},\frac{3}{40},\frac{19}{40},\frac{7}{8}\right\}_B$ & $\left\{\frac{35}{2},\frac{5}{6},\frac{7}{6},\frac{49}{48},\frac{65}{48},\frac{27}{16}\right\}_B$ & $\left\{\frac{465}{14},\frac{25}{14},\frac{27}{14},\frac{31}{16},\frac{249}{112},\frac{265}{112}\right\}_B$\\
        \hline
            \cellcolor{gray!20}$\left\{2,\frac{1}{6},\frac{1}{3},\frac{1}{12},\frac{5}{12},\frac{3}{4}\right\}_S$ $\ast$ & $\left\{\frac{130}{7},\frac{13}{14},\frac{8}{7},\frac{5}{4},\frac{39}{28},\frac{47}{28}\right\}_B$ & 
            \cellcolor{gray!20}$\left\{\frac{182}{5},\frac{11}{10},\frac{11}{5},\frac{49}{20},\frac{57}{20},\frac{13}{4}\right\}_B$\\
        \hline
            $\left\{\frac{18}{5},\frac{3}{10},\frac{2}{5},\frac{3}{20},\frac{11}{20},\frac{3}{4}\right\}_S$ $\ast$ & $\left\{19,\frac{9}{10},\frac{8}{5},\frac{19}{40},\frac{51}{40},\frac{15}{8}\right\}_N$ & \cellcolor{gray!20}$\left\{39,\frac{7}{6},\frac{7}{3},\frac{65}{24},\frac{73}{24},\frac{27}{8}\right\}_B$\\
        \hline
            $\left\{\frac{81}{10},\frac{3}{10},\frac{9}{10},\frac{9}{80},\frac{57}{80},\frac{21}{16}\right\}_N$ & $\left\{\frac{99}{5},\frac{9}{10},\frac{6}{5},\frac{11}{8},\frac{63}{40},\frac{71}{40}\right\}_B$ & $\left\{\frac{391}{10},\frac{13}{10},\frac{12}{5},\frac{207}{80},\frac{239}{80},\frac{51}{16}\right\}_B$\\
        \hline
            $\left\{\frac{42}{5},\frac{3}{5},\frac{7}{10},\frac{7}{20},\frac{3}{4},\frac{19}{20}\right\}_S$ $\ast$ & $\left\{\frac{207}{10},\frac{13}{10},\frac{8}{5},\frac{23}{80},\frac{119}{80},\frac{27}{16}\right\}_N$ & 
            \cellcolor{gray!20}$\left\{\frac{204}{5},\frac{6}{5},\frac{12}{5},3,\frac{16}{5},\frac{17}{5}\right\}_B$\\
        \hline
            $\left\{9,\frac{3}{5},\frac{9}{10},\frac{9}{40},\frac{5}{8},\frac{41}{40}\right\}_N$ & $\left\{21,\frac{7}{6},\frac{4}{3},\frac{9}{8},\frac{35}{24},\frac{43}{24}\right\}_B$ & $\left\{42,\frac{11}{10},\frac{12}{5},\frac{63}{20},\frac{67}{20},\frac{15}{4}\right\}_B$\\
        \hline
            $\left\{10,\frac{2}{3},\frac{5}{6},\frac{5}{12},\frac{3}{4},\frac{13}{12}\right\}_S$ $\ast$ & \cellcolor{gray!20}$\left\{22,\frac{5}{6},\frac{5}{3},\frac{11}{12},\frac{19}{12},\frac{9}{4}\right\}_S$ $\ast$ & $\left\{\frac{297}{7},\frac{19}{14},\frac{18}{7},\frac{165}{56},\frac{181}{56},\frac{27}{8}\right\}_B$\\
        \hline
            $\left\{\frac{143}{14},\frac{11}{14},\frac{13}{14},\frac{39}{112},\frac{55}{112},\frac{17}{16}\right\}_N$ & $\left\{22,\frac{11}{10},\frac{7}{5},\frac{5}{4},\frac{33}{20},\frac{37}{20}\right\}_B$ & \cellcolor{gray!20}$\left\{44,\frac{4}{3},\frac{8}{3},3,\frac{10}{3},\frac{11}{3}\right\}_B$\\
        \hline
            $\left\{\frac{165}{14},\frac{5}{7},\frac{15}{14},\frac{45}{112},\frac{11}{16},\frac{141}{112}\right\}_N$ & $\left\{\frac{221}{10},\frac{13}{10},\frac{7}{5},\frac{17}{16},\frac{117}{80},\frac{149}{80}\right\}_B$ & $\left\{\frac{310}{7},\frac{17}{14},\frac{18}{7},\frac{93}{28},\frac{97}{28},\frac{15}{4}\right\}_B$\\
        \hline
            $\left\{\frac{63}{5},\frac{4}{5},\frac{9}{10},\frac{27}{40},\frac{7}{8},\frac{51}{40}\right\}_B$ & $\left\{\frac{170}{7},\frac{17}{14},\frac{11}{7},\frac{5}{4},\frac{51}{28},\frac{55}{28}\right\}_B$ & 
            \cellcolor{gray!20}$\left\{\frac{228}{5},\frac{7}{5},\frac{14}{5},3,\frac{17}{5},\frac{19}{5}\right\}_B$\\
        \hline
            $\left\{\frac{66}{5},\frac{4}{5},\frac{11}{10},\frac{11}{20},\frac{3}{4},\frac{27}{20}\right\}_S$ $\ast$ & $\left\{\frac{171}{7},\frac{19}{14},\frac{11}{7},\frac{9}{8},\frac{95}{56},\frac{111}{56}\right\}_B$ & $\left\{70,\frac{4}{3},\frac{25}{6},\frac{55}{12},\frac{21}{4},\frac{83}{12}\right\}_B$\\
        \hline
            $\left\{\frac{195}{14},\frac{1}{14},\frac{19}{14},\frac{75}{112},\frac{13}{16},\frac{267}{112}\right\}_B$ & $\left\{\frac{247}{10},\frac{13}{10},\frac{8}{5},\frac{19}{16},\frac{143}{80},\frac{159}{80}\right\}_B$ & $~$\\
        \hline
            $\left\{\frac{195}{14},\frac{6}{7},\frac{15}{14},\frac{75}{112},\frac{13}{16},\frac{155}{112}\right\}_B$ & 
            \cellcolor{gray!20}$\left\{\frac{133}{5},\frac{9}{10},\frac{9}{5},\frac{57}{40},\frac{81}{40},\frac{21}{8}\right\}_B$ & $~$
            \\
        \hline
    \end{tabular}\caption{Rank 3 non-degenerate unitary theories with $\ell = 0$. Type: S=SUSY, B=SUSY Broken, N = non-SUSY.  We highlight the theories that are a product of lower rank theories. Theories with asterisk are discovered in \cite{Bae:2021mej}.}\label{Tab:rank3l0}
\end{table}
and \ref{Tab:rank3l1} respectively. 
\begin{table}[t!]
\centering
     \def\arraystretch{1.7}
     \newcolumntype{?}{!{\vrule width 1.5pt}}
   \begin{tabular}{c|c|c}
      \hline
      \multicolumn{3}{c}{\text{Rank three, $\ell=1$},$\ \{c, h_1^{\text{NS}}, h_2^{\text{NS}}, h_0^{\text{R}}, h_1^{\text{R}},h_2^{\text{R}} \}_{\text{type}}$ }\\
       \hline\hline
$\left\{\frac{13}{7},\frac{1}{7},\frac{3}{14},\frac{5}{56},\frac{3}{8},\frac{29}{56}\right\}_B$ & $\left\{\frac{62}{7},\frac{9}{14},\frac{5}{7},\frac{1}{4},\frac{11}{28},\frac{27}{28}\right\}_N$ & $\left\{\frac{139}{7},\frac{3}{14},\frac{8}{7},\frac{75}{56},\frac{99}{56},\frac{21}{8}\right\}_B$\\\hline
$\left\{2,\frac{1}{5},\frac{3}{10},\frac{1}{20},\frac{1}{4},\frac{9}{20}\right\}_N$ & $\left\{10,\frac{7}{10},\frac{4}{5},\frac{1}{4},\frac{9}{20},\frac{21}{20}\right\}_N$ & $\left\{20,\frac{4}{5},\frac{6}{5},\frac{13}{10},\frac{3}{2},\frac{17}{10}\right\}_B$\\\hline
$\left\{\frac{22}{7},\frac{2}{7},\frac{5}{14},\frac{3}{28},\frac{1}{4},\frac{15}{28}\right\}_N$ & $\left\{\frac{149}{14},\frac{5}{14},\frac{11}{14},\frac{53}{112},\frac{117}{112},\frac{19}{16}\right\}_B$ & $\left\{\frac{146}{7},\frac{8}{7},\frac{17}{14},\frac{25}{28},\frac{41}{28},\frac{7}{4}\right\}_B$\\\hline
$\left\{\frac{89}{14},\frac{3}{14},\frac{9}{14},\frac{9}{112},\frac{73}{112},\frac{15}{16}\right\}_N$ & $\left\{\frac{85}{7},\frac{11}{14},\frac{6}{7},\frac{3}{8},\frac{37}{56},\frac{69}{56}\right\}_N$ & $\left\{\frac{148}{7},\frac{6}{7},\frac{9}{7},\frac{19}{14},\frac{3}{2},\frac{25}{14}\right\}_B$\\\hline
$\left\{7,\frac{3}{10},\frac{7}{10},\frac{3}{40},\frac{27}{40},\frac{7}{8}\right\}_N$ & $\left\{\frac{132}{7},\frac{5}{7},\frac{8}{7},\frac{17}{14},\frac{3}{2},\frac{23}{14}\right\}_B$ & $\left\{\frac{45}{2},\frac{6}{5},\frac{13}{10},\frac{81}{80},\frac{129}{80},\frac{29}{16}\right\}_B$\\\hline

    \end{tabular}\caption{Rank 3 non-degenerate unitary theories with $\ell = 1$. Type: S=SUSY, B=SUSY Broken, N=non-SUSY.}\label{Tab:rank3l1}
\end{table}

As expected, all the non-degenerate SUSY solutions found in \cite{Bae:2021mej} 
are rediscovered in this approach. However we stress that most of the solutions are 
entirely new. One can construct a three-character FRCFT by tensoring 
a two-character FRCFT in Table \ref{tab:rktwo1} with itself.
Some of the theories obtained in this manner are highlighted in the Table \ref{Tab:rank3l0} 
while others from Table \ref{tab:rktwo1} become degenerate. We also note that non-supersymmetric FRCFTs 
start to appear from rank three. 

Some of the theories can be recognized with well-known theories. 
For example, two copies of the $\widehat{su}(2)_3$ WZW model can be 
fermionized to the theory with $c = 18/5$. 
On the other hand, the theory with $c=63/5$ can be bosonized 
to the orbifold theory $\widehat{su}(8)_2/\mathbb{Z}_2$, recently 
studied in \cite{Bae:2021lvk}. 

We remark that families $1$ and $2$ in Table \ref{Tab:exprankthree} of Appendix \ref{App:exponents} 
do not lead to any unitary theories that are physically acceptable.

%%%%%%%%%%%%%%%%%%%%%%%%%%%%%%%%%%%%%%%%%%%%%%%%%%%%%%%%%%%%%%%%%%%%%%%%%%%%%%%%%%%%%%%%%%%%%
%%%%%%%%%%%%%%%%%%%%%%%%%%%%%%%%%%%%%%%%%%%%%%%%%%%%%%%%%%%%%%%%%%%%%%%%%%%%%%%%%%%%%%%%%%%%%

\subsection{Rank 4 with $\ell=0$}

We now proceed to the case of $d=4$. 
The most general rigid fourth-order FMLDE has to be monic, 
which takes the following form 
\begin{align}
    \Bigg[
    \mathcal{D}^4 & + \mu_1(\vartheta_4^4 - \vartheta_2^4)\mathcal{D}^3 + \Big\{ \mu_2 (\vartheta_2^8 + \vartheta_4^8) + \mu_3 \vartheta_2^4 \vartheta_4^4\Big\} \mathcal{D}^2 
    + \Big\{ \mu_4(\vartheta_4^{8}\vartheta_2^4 - \vartheta_2^{8}\vartheta_4^4)
    \\ 
     & + \mu_5 (\vartheta_4^{12} - \vartheta_2^{12}) \Big\} \mathcal{D} + \mu_6 (\vartheta_2^{16} + \vartheta_4^{16}) + \mu_7(\vartheta_4^{12}\vartheta_2^4 + \vartheta_2^{12}\vartheta_4^4) + \mu_8 \vartheta_2^8 \vartheta_4^8 \Bigg]\chi^{\text{NS}}(\tau)=0.
     \nonumber 
\end{align}
The corresponding equation in the R sector is 
\begin{align}
    \Bigg[ 
    \mathcal{D}^4 & - \mu_1(\vartheta_3^4 + \vartheta_4^4)\mathcal{D}^3 + \Big\{ \mu_2 (\vartheta_4^8 + \vartheta_3^8) - \mu_3 \vartheta_4^4 \vartheta_3^4\Big\}\mathcal{D}^2 + \big\{ -\mu_4 (\vartheta_3^{8}\vartheta_4^4 + \vartheta_4^{8}\vartheta_3^4)  
    \\
    & + \mu_5 (\vartheta_3^{12} + \vartheta_4^{12})\Big\} \mathcal{D} 
    + \mu_6 (\vartheta_4^{16} + \vartheta_3^{16}) - \mu_7(\vartheta_3^{12}\vartheta_4^4 + \vartheta_4^{12}\vartheta_3^4) + \mu_8 \vartheta_4^8 \vartheta_3^8 \Bigg]\chi^{\text{R}}(\tau)=0. \nonumber 
\end{align}
Notice that the $d=4$ FMLDE with $\ell=1$ already becomes non-rigid. 

Again, in order to avoid having too many solutions, 
we further impose the unitarity constraint. 
Altogether we have $37$ unitary FRCFTs without free fermion, listed in Table \ref{tab:rk4l0}. 

\begin{table}[t!]
\centering
\scalebox{0.83}{
     \def\arraystretch{1.7}
     \newcolumntype{?}{!{\vrule width 1.5pt}}
   \begin{tabular}{c|c|c}
      \hline
      \multicolumn{3}{c}{\text{Rank four}, $\ell=0$, $\{c, h_1^{\text{NS}}, h_2^{\text{NS}},h_3^{\text{NS}}, h_0^{\text{R}}, 
      h_1^{\text{R}},h_2^{\text{R}},h_3^{\text{R}} \}$ }\\
       \hline\hline
$\left\{\frac{4}{5},\frac{1}{40},\frac{1}{8},\frac{2}{5},\frac{1}{40},\frac{1}{8},\frac{21}{40},\frac{13}{8}\right\}_N$ & $\left\{\frac{63}{8},\frac{3}{8},\frac{5}{8},\frac{3}{4},\frac{21}{64},\frac{49}{64},\frac{69}{64},\frac{81}{64}\right\}_S$ &$\left\{\frac{207}{7},\frac{9}{14},\frac{10}{7},\frac{33}{14},\frac{69}{56},\frac{101}{56},\frac{141}{56},\frac{27}{8}\right\}_S$\\\hline
$\left\{\frac{5}{6},\frac{1}{18},\frac{1}{6},\frac{1}{3},\frac{1}{48},\frac{35}{144},\frac{11}{16},\frac{65}{48}\right\}_N$ & $\left\{\frac{35}{4},\frac{7}{12},\frac{3}{4},\frac{5}{6},\frac{35}{96},\frac{21}{32},\frac{95}{96},\frac{33}{32}\right\}_S$ &$\left\{\frac{247}{8},\frac{7}{8},\frac{5}{4},\frac{17}{8},\frac{113}{64},\frac{133}{64},\frac{181}{64},\frac{209}{64}\right\}_B$\\\hline
$\left\{\frac{7}{8},\frac{1}{8},\frac{1}{4},\frac{7}{8},\frac{1}{64},\frac{5}{64},\frac{21}{64},\frac{33}{64}\right\}_N$ & $\left\{\frac{99}{8},\frac{3}{4},\frac{7}{8},\frac{9}{8},\frac{33}{64},\frac{45}{64},\frac{77}{64},\frac{81}{64}\right\}_S$ &$\left\{\frac{247}{8},\frac{13}{8},\frac{7}{4},\frac{15}{8},\frac{113}{64},\frac{117}{64},\frac{133}{64},\frac{145}{64}\right\}_B$\\\hline
$\left\{1,\frac{1}{32},\frac{1}{8},\frac{9}{32},\frac{1}{32},\frac{9}{32},\frac{25}{32},\frac{49}{32}\right\}_N$ & $\left\{\frac{68}{5},\frac{17}{40},\frac{4}{5},\frac{9}{8},\frac{17}{40},\frac{9}{8},\frac{13}{8},\frac{77}{40}\right\}_N$ &$\left\{\frac{65}{2},\frac{13}{24},\frac{7}{6},\frac{15}{8},\frac{39}{16},\frac{125}{48},\frac{149}{48},\frac{63}{16}\right\}_B$\\\hline
$\left\{1,\frac{1}{14},\frac{2}{7},\frac{9}{14},\frac{1}{56},\frac{9}{56},\frac{25}{56},\frac{7}{8}\right\}_N$ & $\left\{15,\frac{15}{32},\frac{7}{8},\frac{39}{32},\frac{15}{32},\frac{39}{32},\frac{55}{32},\frac{63}{32}\right\}_N$ &$\left\{\frac{663}{20},\frac{17}{10},\frac{7}{4},\frac{39}{20},\frac{65}{32},\frac{69}{32},\frac{357}{160},\frac{377}{160}\right\}_B$\\\hline
$\left\{\frac{5}{4},\frac{1}{32},\frac{5}{32},\frac{1}{4},\frac{1}{16},\frac{5}{16},\frac{9}{16},\frac{29}{16}\right\}_B$ & $\left\{\frac{88}{5},\frac{4}{5},\frac{11}{10},\frac{6}{5},\frac{11}{10},\frac{13}{10},\frac{3}{2},\frac{17}{10}\right\}_B$ &$\left\{\frac{133}{4},\frac{19}{32},\frac{5}{4},\frac{63}{32},\frac{19}{8},\frac{21}{8},\frac{25}{8},\frac{31}{8}\right\}_B$\\\hline
$\left\{\frac{5}{4},\frac{1}{12},\frac{1}{4},\frac{5}{6},\frac{5}{96},\frac{3}{32},\frac{41}{96},\frac{23}{32}\right\}_S$ & $\left\{\frac{56}{3},\frac{7}{9},\frac{7}{6},\frac{4}{3},\frac{7}{6},\frac{23}{18},\frac{3}{2},\frac{11}{6}\right\}_B$ &$\left\{\frac{264}{7},\frac{13}{14},\frac{25}{14},\frac{18}{7},\frac{51}{28},\frac{75}{28},\frac{13}{4},\frac{99}{28}\right\}_B$\\\hline
$\left\{\frac{10}{7},\frac{1}{14},\frac{1}{7},\frac{5}{14},\frac{1}{14},\frac{5}{14},\frac{9}{14},\frac{3}{2}\right\}_B$ & $\left\{\frac{77}{4},\frac{11}{12},\frac{7}{6},\frac{5}{4},\frac{121}{96},\frac{133}{96},\frac{51}{32},\frac{55}{32}\right\}_B$ &$\left\{\frac{196}{5},\frac{63}{80},\frac{11}{10},\frac{35}{16},\frac{49}{16},\frac{253}{80},\frac{293}{80},\frac{73}{16}\right\}_B$\\\hline
$\left\{\frac{3}{2},\frac{1}{18},\frac{2}{9},\frac{8}{9},\frac{11}{144},\frac{3}{16},\frac{59}{144},\frac{107}{144}\right\}_B$  & $\left\{\frac{429}{20},\frac{11}{10},\frac{5}{4},\frac{27}{20},\frac{39}{32},\frac{231}{160},\frac{291}{160},\frac{59}{32}\right\}_B$ &\cellcolor{gray!20} $\left\{\frac{399}{10},\frac{9}{10},\frac{9}{5},\frac{27}{10},\frac{171}{80},\frac{219}{80},\frac{267}{80},\frac{63}{16}\right\}_B$\\\hline
\cellcolor{gray!20}$\left\{\frac{21}{10},\frac{1}{10},\frac{1}{5},\frac{3}{10},\frac{9}{80},\frac{41}{80},\frac{73}{80},\frac{21}{16}\right\}_B$ &$\left\{\frac{70}{3},\frac{7}{6},\frac{14}{9},\frac{11}{6},\frac{7}{12},\frac{5}{4},\frac{65}{36},\frac{23}{12}\right\}_N$ &$\left\{\frac{169}{4},\frac{13}{32},\frac{5}{4},\frac{81}{32},\frac{45}{16},\frac{53}{16},\frac{65}{16},\frac{89}{16}\right\}_B$\\\hline
$\left\{\frac{12}{5},\frac{3}{40},\frac{1}{5},\frac{3}{8},\frac{3}{40},\frac{3}{8},\frac{7}{8},\frac{63}{40}\right\}_{N}$ & $\left\{\frac{203}{8},\frac{7}{8},\frac{13}{8},\frac{9}{4},\frac{29}{64},\frac{81}{64},\frac{125}{64},\frac{161}{64}\right\}_N$ &\cellcolor{gray!20}$\left\{\frac{273}{5},\frac{11}{10},\frac{11}{5},\frac{33}{10},\frac{147}{40},\frac{163}{40},\frac{179}{40},\frac{39}{8}\right\}_B$\\\hline
$\left\{\frac{21}{8},\frac{1}{8},\frac{3}{8},\frac{3}{4},\frac{3}{64},\frac{15}{64},\frac{35}{64},\frac{63}{64}\right\}_N$ & $\left\{27,\frac{13}{14},\frac{12}{7},\frac{33}{14},\frac{27}{56},\frac{75}{56},\frac{115}{56},\frac{21}{8}\right\}_N$ & $\quad$\\\hline
$\left\{\frac{45}{7},\frac{5}{14},\frac{4}{7},\frac{9}{14},\frac{15}{56},\frac{39}{56},\frac{55}{56},\frac{9}{8}\right\}_S$ & $\left\{\frac{225}{8},\frac{5}{8},\frac{11}{8},\frac{9}{4},\frac{75}{64},\frac{111}{64},\frac{155}{64},\frac{207}{64}\right\}_S$ & $\quad$\\\hline

    \end{tabular}}\caption{Rank 4 non-degenerate unitary theories with $\ell = 0$. Type: S=SUSY, B=SUSY Broken, N = non-SUSY. We highlight the theories that 
are a product of lower rank theories.}\label{tab:rk4l0}
\end{table}
Similar to the rank three case, one can construct a four-character FRCFT by tensoring a two-character FRCFT in Table \ref{tab:rktwo1} three times. We highlight those theories that can be obtained in this fashion in Table \ref{tab:rk4l0}. 
Others in Table \ref{tab:rktwo1} generate the degenerate theories with $d=3$. 
What's more, as shown in \cite{Bae:2021lvk}, the $\widehat{su}(4)_3$ WZW model can be mapped to
to the theory with $c=45/7$ via the generalized Jordan-Wigner transformation
while the $\widehat{su}(6)_2$ WZW model to the theory with $c=35/4$. Finally we remark that families $1, 2, 4, 8, 9, 10, 11, 14, 15, 20, 21, 22,$ $25, 27, 28, 29, 30, 32, 33, 39, 42, 43, 44, 46, 47, 48, 49, 52, 53, 54, 55$ in Table \ref{Tab:exprankfour} of Appendix \ref{App:exponents} 
do not lead to any unitary theories with $\ell = 0$ that are physically acceptable.

\subsection*{Acknowledgements}
We would like to thank Nathan Benjamin, Ying-Hsuan Lin and Kaiwen Sun for helpful discussions. We thank Jinbeom Bae and Matthieu Sarkis for comments on the draft. We also thank KIAS Center for Advanced Computation for providing computing resources. ZD thanks LPENS, Paris and LPTHE, Paris for hospitality where parts of this project were carried out. The main result of this work was presented by ZD at KIAS Autumn Symposium on Fields and Strings 2022, and ZD is grateful to the audience for feedback. ZD, KL, SL and LL are supported by KIAS Grant PG076902, PG006904, PG056502 and PG087101 respectively. KL is supported in part by the  National Research Foundation of Korea (NRF) Grant funded by the Korea government (MSIT) (No.2017R1D1A1B06034369).

\newpage

\appendix
\section{Some Finite Group Theory}\label{App:Group}
In this appendix we give the proof of the claim just below eq. \eqref{deneven} in Section \ref{Sec:Integrality}. To set up the notation, we will use $G$ to denote a generic finite group, and $H$ for one of its subgroup, $|G|,|H|$ for their orders, $|G/H|$ for the corresponding index. 

For application in this paper, we can take  $G$ as $	\mathrm{SL}\left(2, \mathbb{Z}_{N}\right)  $, and $H$ as one of  $\G_\th / \Gamma(N)$, $\G_0(2) / \Gamma(N)$,$\G^0(2) / \Gamma(N)$. The order of $G$ is
\begin{equation}\label{order SL}
|	\mathrm{SL}\left(2, \mathbb{Z}_{N}\right) | = N^3 \prod_{p| N} (1 - \frac{1}{p^2})
\end{equation}
 and we have  $|G/H|=3$ for all the three choices of $H$.  We are mainly interested in the special case $N=2^{\lambda}$, here the corresponding $H$ is also denoted as $P_{\lambda}$.

Suppose $V$ is a finite dimensional vector space over some algebraic closed field $K$ with $\operatorname{Char}K=0$ \footnote{Essentially, we have $K=\mathbb{C}$ in this paper.}. Having a representation $\rho: G\rightarrow \mathrm{GL}(V)$ of $G$ on $V$ is equivalent to assigning a $K[G]$-module structure for $V$, so we can just use the $K[G]$-module $V$ to refer to this representation. If we restrict this representation to some subgroup $H$, we will get a representation $\rho|_H: H\rightarrow \mathrm{GL}(W)$ of $H$ on some subspace $W$ of $V$, and we say $W$ is the restriction of $V$ on $H$. Alternatively, starting with some $W$, that is, a representation $\xi: H\rightarrow \mathrm{GL}(W)$, we can construct a 
representation $W'$ of $G$ through the following scalar extension
\begin{equation}
    W'=K[G]\bigotimes_{\quad K[H]}W.
\end{equation}
We say $V$ is induced by $W$ iff $V\cong W'$ as $K[G]$ module. More concretely, this means

\begin{equation}
	V=\bigoplus_{\sigma\in  G/H} W_{\sigma},\quad  W_{\sigma}=\rho(s_{\sigma})W.
\end{equation}
Then $ H$ is precisely the collection of elements $h\in G$ such that $\rho(h)=\xi(h),\, \rho(h)W=W$.
In particular, when $H$ is a Sylow p-subgroup and $V,W$ are irreducible, we say $V$ is p-induced by $W$.

It is obvious that
\begin{equation}\label{dim induced}
	\dim V= |G/H| \dim W.
\end{equation}

If we begin with some irreducible $V$, its restriction $W$ may or may not be irreducible. On the other hand, if we begin with some irreducible $W$, the induced $W'$ may or may not be irreducible as well, but at least we have:
\begin{thm}
Every irreducible $V$ is contained in some induced  $W'$ by some irreducible $W$.
\end{thm}
When the induced representation $W'$ itself is irreducible, $W$ must be irreducible as well, then we may use eq. \eqref{dim induced} to calculate the dimension of $W$ if we know the dimension of $V\cong W'$.

For example, we can take $G=\operatorname{SL}(2, \mathbb{Z}_{2^{\lambda}})$ and $H=P_{\lambda}$ with $|G/H|=3$. In table 1 of \cite{Kaidi:2021ent}, we find when $\lambda>5$ the irreducible representations have dimensions:

\begin{equation}
	\left\{2^{\lambda-1}\right\} \cup\left\{3 \cdot 2^{\lambda-i} \mid i=1,2,3,4\right\}.
\end{equation}

So it seems that the $3 \cdot 2^{\lambda-i}$ type representations are 2-induced. Usually it is hard to decide whether a restricted or induced representation is irreducible or not in general, but in this case we have the following theorem \cite{RIESE1998682}, which in our notation says

\begin{thm}
Given a prime $p$ dividing $|G|$, an irreducible $V$ is p-induced iff $|G|/\operatorname{dim}V=p^r$ for some positive integer $r$.
\end{thm}

Applying this theorem to our problem with $p=2$ and using eq. \eqref{order SL} we find $|G|=3\cdot 2^{m(\lambda)}$ for some positive integer $m(\lambda)>\lambda$, so when $\lambda >5$ we have the following set of $2$-induced representations of $G=\operatorname{SL}(2, \mathbb{Z}_{2^{\lambda}})$

\begin{equation}
	\left\{3 \cdot 2^{\lambda-i} \mid i=1,2,3,4\right\}.
\end{equation}

Then from $|G/H|=3$ we find the set of irreducible representations of $H=P_{\lambda}$ labeled by their dimensions
\begin{equation}
	\left\{  2^{\lambda-i} \mid i=1,2,3,4\right\}.
\end{equation}

While the $\lambda\leq 5$ cases can be treated by this method or calculated directly, then we can construct a fermionic version of Table 1 of \cite{Kaidi:2021ent} .

\section{Exponents}\label{App:exponents}
In this appendix, we collect the exponents for rank $d = 2, 3, 4$ in Tables \ref{Tab:expranktwo}, \ref{Tab:exprankthree}, \ref{Tab:exprankfour}. Recall that they always appear in terms of families, so we only need to give one representative for each family.
\begin{table}[t!]
\centering
     \def\arraystretch{1.7}
     \newcolumntype{?}{!{\vrule width 1.5pt}}   
\begin{tabular}{c?c||c?c}
      \hline
       \multicolumn{4}{c}{$d = 2$ \text{exponents}: 
       $\{2\alpha^{\text{NS}},\alpha^{\text{R} }\} + \frac{k}{24} \{-{\bf 1}_2 ,{\bf 1}_2\} \text{ for } k\in\mathbb{Z}$}\\
    \hline\hline
        \text{No}. & 
        \text{Representative} &
        \text{No}. & 
        \text{Representative}
        \\
    \hline
        1& $\{(0,\frac{1}{8}), (\frac{7}{16} , \frac{15}{16} ) \}$ & 
        8& $\{(0, \frac{1}{3} ), (\frac{1}{4} , \frac{11}{12} )\}$\\
    \hline
        2& $\{(0 ,\frac{1}{4}  ), (\frac{3}{8}  , \frac{7}{8}  ) \}$ &
        9& $\{(0 , \frac{1}{3} ), (\frac{5}{12} , \frac{3}{4} )\}$
        \\
    \hline
        3&$\{(0 ,\frac{3}{8}  ), (\frac{5}{16} , \frac{13}{16} )\}$ &
        10&$\{(\frac{1}{120} , \frac{49}{120} ), (\frac{17}{120} , \frac{113}{120} )\}$ 
        \\
    \hline
        4&$\{(0, \frac{1}{2} ), (\frac{1}{8} , \frac{7}{8} )\}$ & 
        11&$\{(\frac{1}{120} , \frac{49}{120} ), (\frac{53}{120} , \frac{77}{120} )\}$ 
        \\
    \hline
        5&$\{(0 , \frac{1}{2} ), (\frac{1}{4} , \frac{3}{4} )\}$ & 
        12&$\{(\frac{1}{60} , \frac{49}{60} ), (\frac{1}{30} , \frac{19}{30} )\}$
        \\
    \hline
        6&$\{(\frac{1}{48} , \frac{25}{48} ), (\frac{1}{24} , \frac{11}{12} )\}$ & 
        13&$\{(\frac{1}{60} , \frac{49}{60} ), (\frac{2}{15} , \frac{8}{15} )\}$ 
        \\
    \hline
        7&$\{(\frac{1}{48} , \frac{25}{48} ), (\frac{1}{6} , \frac{19}{24} )\}$ &  &
        \\
    \hline
\end{tabular}
    \caption{All possible exponents mod 1 for rank 2 non-degenerate FRCFTs.}
    \label{Tab:expranktwo}
\end{table}
\begin{table}[t!]
\centering
     \def\arraystretch{1.7}
     \newcolumntype{?}{!{\vrule width 1.5pt}}   
\begin{tabular}{c?c||c?c}
      \hline
       \multicolumn{4}{c}{$d = 3$ \text{exponents}: 
       $\{2\alpha^{\text{NS}},\alpha^{\text{R} }\} + \frac{k}{24} \{-{\bf 1}_3 ,{\bf 1}_3\} \text{ for } k\in\mathbb{Z}$}\\
    \hline\hline
        \text{No}. & 
        \text{Representative} &
        \text{No}. & 
        \text{Representative}
        \\
    \hline
        1& $ \{(0,\frac{1}{5},\frac{4}{5}), (0,\frac{2}{5} , \frac{3}{5} ) \}$ & 
        6& $\{(0,\frac{1}{3}, \frac{2}{3}), (0,\frac{1}{3} , \frac{2}{3} )\}$
        \\
    \hline
        2& $ \{(0,\frac{1}{5},\frac{4}{5}), (\frac{1}{10},\frac{1}{2}  , \frac{9}{10}  ) \}$ &
        7& $\{(\frac{1}{168}, \frac{25}{168}, \frac{121}{168}), (\frac{11}{168} ,\frac{107}{168}, \frac{155}{168} )\}$
        \\
    \hline
        3&$\{(0,\frac{2}{5}, \frac{3}{5}), (0,\frac{1}{5},\frac{4}{5} )\}$ &
        8&$\{(\frac{1}{168}, \frac{25}{168}, \frac{121}{168}), (\frac{23}{168} , \frac{71}{168},\frac{95}{168} )\}$ 
        \\
    \hline
        4&$\{(0,\frac{2}{5}, \frac{3}{5}), (\frac{3}{10},\frac{1}{2},\frac{7}{10} )\}$ & 
        9&$\{(\frac{1}{56}, \frac{9}{56}, \frac{25}{56}), (\frac{11}{56} , \frac{43}{56},\frac{51}{56} )\}$ 
        \\
    \hline
        5&$\{(0,\frac{1}{3}, \frac{2}{3}), (\frac{1}{6} , \frac{1}{2}, \frac{5}{6} ) \}$ & 
        10&$\{(\frac{1}{56}, \frac{9}{56}, \frac{25}{56}), (\frac{15}{56} ,\frac{23}{56} ,\frac{39}{56} )\}$
        \\
    \hline
\end{tabular}
    \caption{All possible exponents mod 1 for rank 3 non-degenerate FRCFTs.}
    \label{Tab:exprankthree}
\end{table}

\begin{table}[t!]
\centering
\scalebox{0.8}{
     \def\arraystretch{1.7}
     \newcolumntype{?}{!{\vrule width 1.5pt}}   
\begin{tabular}{c?c||c?c}
      \hline
       \multicolumn{4}{c}{$d = 4$ \text{exponents}: 
       $\{2\alpha^{\text{NS}},\alpha^{\text{R} }\} + \frac{k}{24} \{-{\bf 1}_4 ,{\bf 1}_4\} \text{ for } k\in\mathbb{Z}$}\\
    \hline\hline
        \text{No}. & 
        \text{Representative} &
        \text{No}. & 
        \text{Representative}
        \\
    \hline
        1& $\{(\frac{1}{5},\frac{2}{5},\frac{3}{5},\frac{4}{5}), (\frac{1}{5},\frac{2}{5},\frac{3}{5},\frac{4}{5} ) \}$ & 
        29 & $\{(\frac{1}{120},\frac{3}{40},\frac{49}{120},\frac{27}{40}),(\frac{23}{120},\frac{47}{120},\frac{21}{40},\frac{29}{40})\}$\\
    \hline
        2& $\{(0,\frac{1}{24},\frac{3}{8},\frac{2}{3}), (\frac{1}{16},\frac{19}{48},\frac{9}{16},\frac{43}{48} ) \}$ & 
        30 & $\{(\frac{1}{120},\frac{2}{15},\frac{49}{120},\frac{8}{15}),(\frac{19}{240},\frac{91}{240},\frac{139}{240},\frac{211}{240})\}$ \\
    \hline
        3& $\{(0,\frac{1}{16},\frac{1}{4},\frac{9}{16}), (\frac{5}{32},\frac{13}{32},\frac{21}{32},\frac{29}{32} )\}$ &
        31 & $\{(\frac{1}{120},\frac{19}{120},\frac{49}{120},\frac{91}{120}),(\frac{1}{15},\frac{4}{15},\frac{17}{30},\frac{23}{30})\}$ \\
    \hline
        4 & $\{(0,\frac{1}{16},\frac{9}{16},\frac{3}{4}), (\frac{1}{32},\frac{9}{32},\frac{17}{32},\frac{25}{32})\}$ & 
        32& $\{(\frac{1}{120},\frac{49}{120},\frac{61}{120},\frac{109}{120}),(\frac{1}{60},\frac{1}{15},\frac{4}{15},\frac{49}{60})\}$ \\
    \hline
        5& $\{(0,\frac{1}{12},\frac{1}{3},\frac{3}{4}), (\frac{1}{24},\frac{3}{8},\frac{13}{24},\frac{7}{8} )\}$ & 
        33 & $\{(\frac{1}{120},\frac{49}{120},\frac{61}{120},\frac{109}{120}),(\frac{23}{120},\frac{47}{120},\frac{83}{120},\frac{107}{120})\}$ \\
    \hline
        6& $\{(0,\frac{1}{9},\frac{4}{9},\frac{7}{9}), (0,\frac{2}{9},\frac{5}{9},\frac{8}{9} )\}$ & 
        34& $\{(\frac{1}{120},\frac{49}{120},\frac{73}{120},\frac{97}{120}),(\frac{11}{120},\frac{59}{120},\frac{83}{120},\frac{107}{120})\}$ \\
    \hline
        7& $\{(0,\frac{1}{9},\frac{4}{9},\frac{7}{9}), (\frac{1}{18},\frac{7}{18},\frac{1}{2},\frac{13}{18} )\}$ & 
        35& $\{(\frac{1}{120},\frac{49}{120},\frac{73}{120},\frac{97}{120}),(\frac{17}{120},\frac{41}{120},\frac{89}{120},\frac{113}{120})\}$ \\
    \hline
        8& $\{(0,\frac{1}{9},\frac{4}{9},\frac{7}{9}), (\frac{5}{36},\frac{1}{4},\frac{17}{36},\frac{29}{36} )\}$ & 
        36& $\{(\frac{1}{120},\frac{49}{120},\frac{73}{120},\frac{97}{120}),(\frac{29}{120},\frac{53}{120},\frac{77}{120},\frac{101}{120})\}$ \\
    \hline
        9& $\{(0,\frac{1}{9},\frac{4}{9},\frac{7}{9}), (\frac{11}{36},\frac{23}{36},\frac{3}{4},\frac{35}{36} )\}$ & 
        37& $\{(\frac{1}{96},\frac{25}{96},\frac{49}{96},\frac{73}{96}),(\frac{1}{48},\frac{1}{12},\frac{25}{48},\frac{5}{6})\}$ \\
    \hline
        10& $\{(0,\frac{1}{8},\frac{1}{3},\frac{11}{24}), (\frac{3}{16},\frac{17}{48},\frac{11}{16},\frac{41}{48} )\}$ &
        38 & $\{(\frac{1}{96},\frac{25}{96},\frac{49}{96},\frac{73}{96}),(\frac{7}{48},\frac{5}{24},\frac{11}{24},\frac{31}{48})\}$ \\
    \hline
        11& $\{(0,\frac{1}{8},\frac{1}{2},\frac{5}{8}), (\frac{1}{16},\frac{5}{16},\frac{9}{16},\frac{13}{16} ) \}$ &  
        39& $\{(\frac{1}{80},\frac{9}{80},\frac{41}{80},\frac{49}{80}),(\frac{1}{40},\frac{9}{40},\frac{13}{20},\frac{17}{20})\}$ \\
    \hline
        12& $\{(0,\frac{1}{7},\frac{2}{7},\frac{4}{7}), (\frac{1}{28},\frac{9}{28},\frac{3}{4},\frac{25}{28} ) \}$ & 
        40& $\{(\frac{1}{80},\frac{9}{80},\frac{41}{80},\frac{49}{80}),(\frac{1}{10},\frac{31}{40},\frac{9}{10},\frac{39}{40})\}$ \\
    \hline
        13& $\{(0,\frac{1}{7},\frac{2}{7},\frac{4}{7}), (\frac{1}{4},\frac{11}{28},\frac{15}{28},\frac{23}{28} )\}$ & 
        41& $\{(\frac{1}{60},\frac{1}{15},\frac{4}{15},\frac{49}{60}),(\frac{1}{120},\frac{49}{120},\frac{61}{120},\frac{109}{120})\}$ \\ 
    \hline
        14& $\{(0,\frac{1}{6},\frac{1}{2},\frac{2}{3}), (0,\frac{1}{3},\frac{1}{2},\frac{5}{6})\}$ & 
        42 & $\{(\frac{1}{60},\frac{17}{120},\frac{49}{60},\frac{113}{120}),(\frac{17}{240},\frac{113}{240},\frac{137}{240},\frac{233}{240})\}$ \\
    \hline
        15& $\{(0,\frac{1}{6},\frac{1}{2},\frac{2}{3}), (\frac{1}{8},\frac{3}{8},\frac{11}{24},\frac{17}{24} )\}$ & 
        43& $\{(\frac{1}{60},\frac{3}{20},\frac{7}{20},\frac{49}{60}),(\frac{1}{20},\frac{23}{60},\frac{9}{20},\frac{47}{60})\}$\\
    \hline
        16& $\{(0,\frac{3}{16},\frac{1}{4},\frac{11}{16}), (\frac{3}{32},\frac{11}{32},\frac{19}{32},\frac{27}{32} )\}$ & 
        44 & $\{(\frac{1}{60},\frac{3}{20},\frac{7}{20},\frac{49}{60}),(\frac{17}{60},\frac{11}{20},\frac{53}{60},\frac{19}{20})\}$ \\
    \hline
        17& $\{(0,\frac{3}{16},\frac{11}{16},\frac{3}{4}), (\frac{7}{32},\frac{15}{32},\frac{23}{32},\frac{31}{32} )\}$ & 
        45  & $\{(\frac{1}{60},\frac{23}{120},\frac{47}{120},\frac{49}{60}),(\frac{83}{240},\frac{107}{240},\frac{203}{240},\frac{227}{240})\}$ \\
    \hline
        18& $\{(0,\frac{2}{9},\frac{5}{9},\frac{8}{9}), (0,\frac{1}{9},\frac{4}{9},\frac{7}{9} )\}$ & 
        46 & $\{(\frac{1}{60},\frac{19}{60},\frac{31}{60},\frac{49}{60}),(\frac{1}{120},\frac{31}{120},\frac{49}{120},\frac{79}{120})\}$\\
    \hline
        19& $\{(0,\frac{2}{9},\frac{5}{9},\frac{8}{9}), (\frac{1}{36},\frac{1}{4},\frac{13}{36},\frac{25}{36} )\}$ & 
        47& $\{(\frac{1}{60},\frac{19}{60},\frac{31}{60},\frac{49}{60}),(\frac{17}{60},\frac{23}{60},\frac{47}{60},\frac{53}{60})\}$\\
    \hline
        20& $\{(0,\frac{2}{9},\frac{5}{9},\frac{8}{9}), (\frac{7}{36},\frac{19}{36},\frac{3}{4},\frac{31}{36} )\}$ & 
        48& $\{(\frac{1}{48},\frac{7}{48},\frac{25}{48},\frac{31}{48}),(\frac{5}{48},\frac{11}{48},\frac{29}{48},\frac{35}{48})\}$\\
    \hline
        21& $\{(0,\frac{2}{9},\frac{5}{9},\frac{8}{9}), (\frac{5}{18},\frac{1}{2},\frac{11}{18},\frac{17}{18} ) \}$ & 
        49 & $\{(\frac{1}{48},\frac{7}{48},\frac{25}{48},\frac{31}{48}),(\frac{17}{48},\frac{23}{48},\frac{41}{48},\frac{47}{48})\}$ \\
    \hline
        22& $\{(0,\frac{1}{4},\frac{1}{2},\frac{3}{4}), (0,\frac{1}{4},\frac{1}{2},\frac{3}{4} ) \}$ &
        50 & $\{(\frac{1}{48},\frac{3}{16},\frac{25}{48},\frac{11}{16}),(0,\frac{1}{8},\frac{2}{3},\frac{19}{24})\}$ \\
    \hline
        23& $\{(0,\frac{3}{7},\frac{5}{7},\frac{6}{7}), (\frac{3}{28},\frac{1}{4},\frac{19}{28},\frac{27}{28} )\}$ & 
        51& $\{(\frac{1}{48},\frac{3}{16},\frac{25}{48},\frac{11}{16}),(\frac{1}{24},\frac{3}{8},\frac{5}{12},\frac{3}{4})\}$\\
    \hline
        24&$\{(0,\frac{3}{7},\frac{5}{7},\frac{6}{7}), (\frac{5}{28},\frac{13}{28},\frac{17}{28},\frac{3}{4})\}$ & 52&$\{(\frac{1}{48},\frac{13}{48},\frac{25}{48},\frac{37}{48}),(\frac{1}{24},\frac{1}{6},\frac{13}{24},\frac{2}{3})\}$\\
    \hline
        25&$\{(\frac{1}{240},\frac{49}{240},\frac{121}{240},\frac{169}{240}), (\frac{1}{120},\frac{17}{60},\frac{49}{120},\frac{53}{60} )\}$ & 
        53 &$\{(\frac{1}{48},\frac{13}{48},\frac{25}{48},\frac{37}{48}),(\frac{5}{48},\frac{17}{48},\frac{29}{48},\frac{41}{48})\}$\\
    \hline
        26& $\{(\frac{1}{240},\frac{49}{240},\frac{121}{240},\frac{169}{240}), (\frac{1}{30},\frac{31}{120},\frac{19}{30},\frac{79}{120})\}$ & 
        54&$\{(\frac{1}{48},\frac{13}{48},\frac{25}{48},\frac{37}{48}),(\frac{11}{48},\frac{23}{48},\frac{35}{48},\frac{47}{48})\}$\\
    \hline
        27&$\{(\frac{1}{120},\frac{1}{30},\frac{49}{120},\frac{19}{30}), (\frac{31}{240},\frac{79}{240},\frac{151}{240},\frac{199}{240} )\}$ & 
        55 &$\{(\frac{1}{32},\frac{9}{32},\frac{17}{32},\frac{25}{32}),(0,\frac{1}{16},\frac{9}{16},\frac{3}{4})\}$\\
    \hline
        28&$\{(\frac{1}{120},\frac{3}{40},\frac{49}{120},\frac{27}{40}), (\frac{1}{40},\frac{9}{40},\frac{83}{120},\frac{107}{120} )\}$ & 
        56 &$\{(\frac{1}{32},\frac{9}{32},\frac{17}{32},\frac{25}{32}),(\frac{1}{8},\frac{3}{16},\frac{3}{8},\frac{11}{16})\}$\\
    \hline
\end{tabular}}
    \caption{All possible exponents mod 1 for rank 4 non-degenerate FRCFTs}
    \label{Tab:exprankfour}
\end{table} 

\clearpage    
\bibliographystyle{JHEP}
\bibliography{main}

\end{document}